\documentclass[12pt, draftclsnofoot, onecolumn]{IEEEtran}
\usepackage{graphicx}
\usepackage{amsmath,bbm,epsfig,amssymb,amsfonts, amstext, verbatim,amsopn,multirow,multicol,lipsum}
\usepackage{cite}
\usepackage[font=scriptsize]{caption}
\usepackage{subcaption}
\usepackage{balance}
\usepackage{url}
\usepackage{amsfonts}
\usepackage{epsfig}
\usepackage{epstopdf}
\usepackage{setspace}
\usepackage{stmaryrd}
\usepackage{psfrag}	
\usepackage{multirow}
\usepackage{float}
\usepackage{cases}  
\usepackage[process=auto]{pstool}
\usepackage{etoolbox}
\usepackage{algorithm}
\usepackage{algorithmic}
\usepackage{hyperref}
\usepackage{wasysym}[nointegrals]
\usepackage{pifont}
\allowdisplaybreaks
\usepackage{graphicx}
\usepackage{wrapfig}
\usepackage{lscape}
\usepackage{rotating}
\usepackage{epstopdf}

\newtheorem{theo}{Theorem}
\newtheorem{lem}{Lemma}

\newtheorem{prop}{Proposition}
\newtheorem{corol}{Corollary}

\newcommand{\pushright}[1]{\ifmeasuring@#1\else\omit\hfill$\displaystyle#1$\fi\ignorespaces}
\newcommand{\pushleft}[1]{\ifmeasuring@#1\else\omit$\displaystyle#1$\hfill\fi\ignorespaces}
\newtoggle{Conf}

\togglefalse{Conf}

\iftoggle{Conf}{%
}{%
}

\EndPreamble

\begin{document}

\title{    Modeling and Design of  IRS-Assisted  Multi-Link FSO Systems 
\vspace{-0.3cm}\footnote{This paper was presented in part at IEEE WCNC 2021\cite{WCNC}.}}
\author{Hedieh Ajam, Marzieh Najafi, Vahid Jamali, Bernhard Schmauss,  and Robert Schober\\  
Friedrich-Alexander University  Erlangen-Nuremberg, Germany
\vspace{-0.2cm}}
\maketitle
\begin{abstract}
In this paper, we  investigate  the modeling and design of  intelligent reflecting surface (IRS)-assisted optical communication systems  which are deployed to relax the line-of-sight (LOS) requirement in multi-link free space optical (FSO) systems. The  FSO  laser beams incident on the optical IRSs have a Gaussian  power intensity profile  and a nonlinear phase profile, whereas the plane waves in radio frequency (RF)  systems have a uniform power intensity profile and a linear phase profile.  Given these substantial differences, the results available for IRS-assisted RF systems are not applicable to IRS-assisted FSO systems.  Therefore, we develop a new analytical channel model for   point-to-point  IRS-assisted  FSO systems  based on the Huygens-Fresnel principle.  Our analytical  model captures the impact of the  size, position, and orientation of the IRS as well as  its phase shift profile on the end-to-end channel.  To allow the sharing of the  optical IRS by multiple FSO links,  we propose three different  protocols, namely the  time division (TD), IRS-division (IRSD), and IRS homogenization (IRSH) protocols.  The proposed protocols address the specific characteristics  of  FSO systems including the non-uniformity and  possible misalignment  of the laser beams.  Furthermore, to compare the proposed IRS sharing protocols, we analyze the bit error rate (BER) and the outage probability  of   IRS-assisted multi-link FSO systems   in the  presence of  inter-link interference. Our simulation results validate the accuracy of the proposed analytical channel model for  IRS-assisted FSO  systems and confirm that this model  is  applicable for both large and intermediate IRS-receiver lens distances.  Furthermore, we show that for the proposed IRSD and IRSH protocols, inter-link interference becomes negligible if the laser  beams are properly centered on the IRS and  the  transceivers are carefully positioned, respectively. Moreover, in the absence of misalignment errors,   the IRSD protocol outperforms the other protocols, whereas in the  presence of misalignment  errors, the IRSH  protocol performs significantly better than the IRSD protocol. 
\end{abstract}
\section{Introduction}
Free space optical (FSO) systems are prime candidates for facilitating the bandwidth-hungry services  of the  next generation of wireless communication networks and beyond \cite{6G}. Due to their directional narrow laser beams, easy-to-install and cost-efficient transceivers, license-free bandwidth, and high data rate,  FSO links are appealing for last-mile access, fiber backup, and backhaul of wireless networks. However, FSO systems require line-of-sight (LOS) link connections and  they are impaired by atmospheric turbulence, beam divergence, and misalignment errors in long-distance deployments. To mitigate these performance-limiting factors, various methods including diversity techniques \cite{diversity}, serial and parallel FSO relays \cite{Safari_relay}, and RF backup links \cite{Marzieh_RFFSO} have been proposed. Recently,  the authors of \cite{Marzieh_IRS, Marzieh_IRS_jou}  proposed the application of optical intelligent reflecting surfaces (IRSs) to  connect  a transmitter  with an obstructed receiver. 

Optical IRSs are planar structures   which  can manipulate the  properties of an incident wave such as its polarization, phase, and amplitude in reflection  and  transmission \cite{Arbabi,Marco_survey}. They can be realized using different technologies including mirrors, micro-mirrors, and metamaterials \cite{Vahid_Magazine}. Mirrors and micro-mirrors only support specular reflection by mechanical adjustment of their orientation. Metamaterials consist of nano-structured antennas, referred to as unit cells, which can resonate in scales much smaller than the wavelength  \cite{ Vahid_Magazine}. Thus,    optical metamaterial-based IRSs can   provide a better control of the spatial resolution of the reflected wave compared to mirrors and micro-mirrors.  In particular, they can apply an abrupt phase shift to the incident wave, changing the accumulated phase of the   wave  and focusing or redirecting the wave in a desired direction.

In radio frequency (RF) wireless  communication systems, IRSs have been exploited to increase coverage, ensure security, harness interference, and  improve the quality of  non-line-of-sight (NLOS) connections \cite{zhang}. Unlike  RF systems,  where the wavefront  incident on the IRS can be modeled as planar and the power is uniformly distributed across the IRS, FSO systems employ    Gaussian laser beams, which have a curved wavefront and a non-uniform power distribution.  Therefore,  a careful study of IRS-assisted FSO systems is needed as existing results from IRS-assisted RF systems are not applicable.  The authors of \cite{Marzieh_IRS,Marzieh_IRS_jou} exploited geometric optics to determine the impact of an IRS on the performance of  an FSO link but ignored  the impact of the IRS size and the lens size in their three-dimensional model. They also employed an equivalent mirror-based analysis to determine the  IRS phase shift profile required for anomalous reflection.  In this paper, we  exploit the Huygens-Fresnel principle  to model the IRS-assisted FSO channel which explicitly captures the impact of the IRS phase shift profile, the IRS size, and lens size. Moreover, the authors of  \cite{practicalIRS} applied an optical IRS to enhance  indoor communication links. In \cite{Alouini_IRS},  the impact of  IRSs on   visible light communication (VLC) was investigated. However,   for VLC,   non-directional beams are employed, which exhibit a  different characteristic compared to the   Gaussian laser beams used in  FSO systems.  

\begin{figure}
	\centering
	\includegraphics[width=0.55\textwidth]{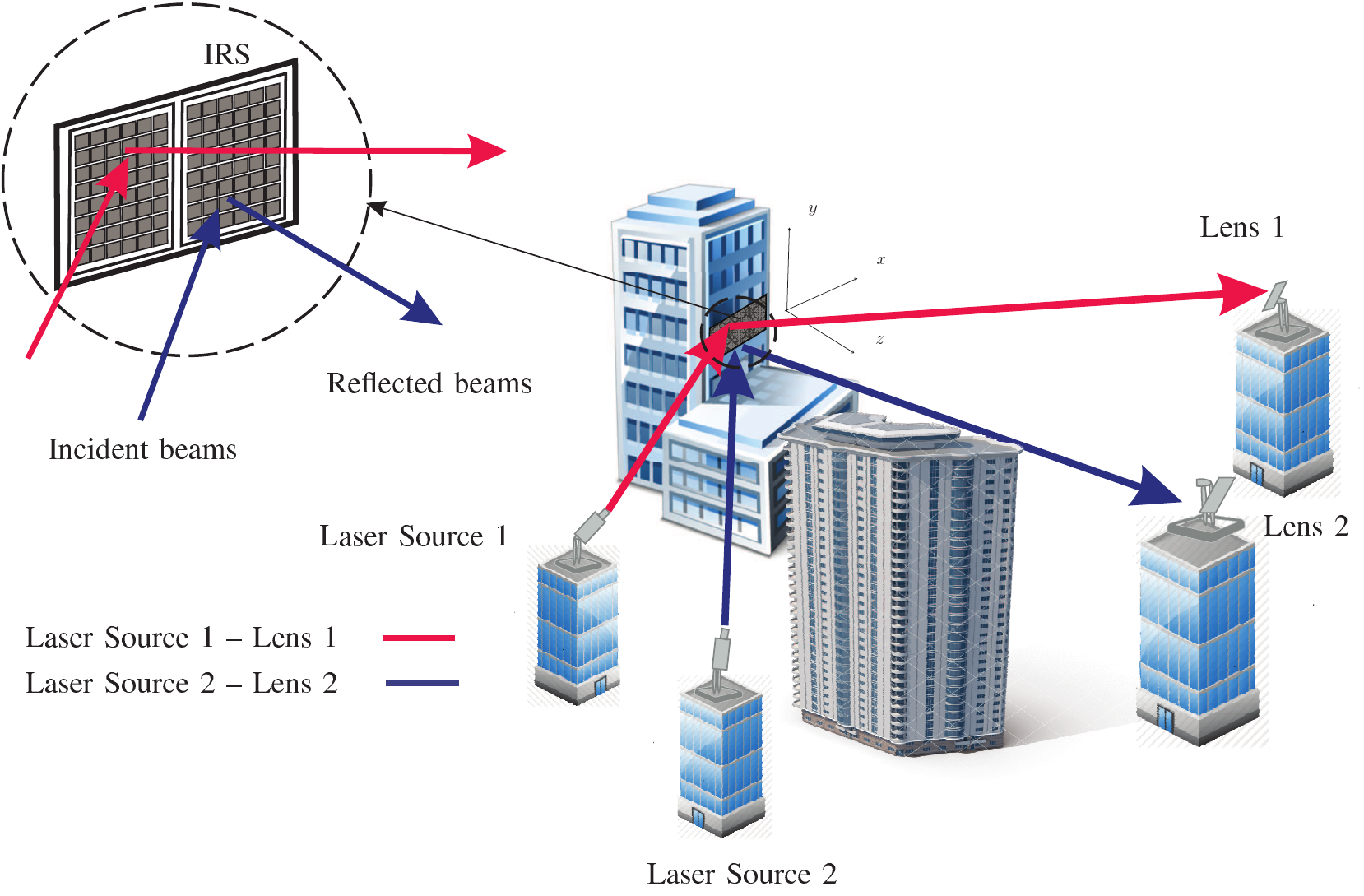}
	\caption{Multiple FSO links sharing a single  IRS.}
	\label{Fig:MA}\vspace*{-6mm}
\end{figure}
Given the Gaussian FSO beams and the flexibility of the IRS, a single IRS surface can be shared among multiple FSO links.   In \cite{MA_reflectors}, the authors investigated  point-to-multi-point FSO communications where a laser source transmits to multiple receivers based on a time-division protocol  using multiple fixed reflectors and a single rotating reflector, respectively. The authors in \cite{Multicast} considered multi-cast transmission where a laser beam illuminates an optical IRS which splits the beam  among multiple FSO receivers. They modeled the reflected power density using Fraunhofer diffraction and far-field approximations. However, we show that the far-field approximation is only valid for specific IRS-receiver lens distances, incident beam widths, and IRS tile sizes.
  
In this paper, we employ  an  IRS-assisted  FSO system to provide a connection between multiple transmitters and their respective obstructed  receivers, see Fig.~\ref{Fig:MA}. The transmitters are equipped with  laser sources (LSs) emitting  Gaussian  beams which are reflected by an IRS towards  respective receivers  where the beams are focused by a lens onto a photo detector (PD). In this work, first, we develop an analytical channel model for  IRS-assisted point-to-point FSO systems, then, we propose three protocols to enable the sharing of the IRS by  multiple  FSO links and analyze the resulting performance in terms of the bit error rate (BER) and outage probability.     In the following, we summarize the main contributions of this work.
\begin{itemize}
\item  Based on the Huygens-Fresnel principle, we analyze the deterministic channel gain  of  a point-to-point  IRS-assisted FSO system   which employs a Gaussian beam  that is emitted by an LS and  reflected by   an optical IRS. The proposed analytical point-to-point channel model takes into account the non-uniform  power distribution and the nonlinear phase profile  of  the Gaussian beam as well as  the  impact of   the relative position of the IRS with respect to (w.r.t.)  the lens and the LS,   the size of the IRS, and the phase shift profile of the IRS.   
\item We show that depending on the IRS size and the width of the incident beam, existing end-to-end channel models based on the far-field approximation  may not always be valid.  We mathematically characterize  the  range of intermediate and far-field distances and	propose an analytical channel  model that is valid for these distances. 
\item Three different  protocols are proposed  to share a single IRS in a multi-link FSO system, namely, protocols based on time division (TD), IRS division (IRSD), and IRS homogenization (IRSH). For each protocol,  the size and  phase shift profile of the IRS, laser beam footprints, and the lens centers on the IRS are specified.  
\item We  consider both linear phase shift (LP) and quadratic phase shift (QP) profiles  across the IRS. The former was considered before in  \cite{Vahid_IRS,Estakhri,Kochkina} and can only change the direction of the reflected beam. The latter is considered for the first time and  can reduce the beam divergence along the propagation path.
\item The performance of the considered multi-link FSO systems is analyzed  in terms of BER and  outage probability. Given that the IRS is shared  by several LS-PD pairs, the inter-link interference potentially caused by the proposed IRS sharing protocols is taken into account  in our analysis. Using simulations, we show that by appropriately choosing the   LS-IRS and IRS-receiver lens distances as well as  the separation angle between the LSs and the receiver lenses, respectively, inter-link  interference can be avoided.  
\item Simulation results confirm our analysis of the deterministic  gain of the IRS-assisted FSO channel. In particular, for large IRS sizes, large  beam widths, and practical IRS-receiver lens distances, the far-field approximation may not be valid whereas our proposed model yields accurate result.
\item Our simulation results also validate the analytical expressions derived for the BER and outage probability.  Our results suggest that the IRSH protocol is preferable in the presence of  misalignment  errors. However, in the absence of misalignment errors,  the IRSD protocol is advantageous as it yields a higher received power than the IRSH protocol  and a lower delay compared to the TD protocol. 
\end{itemize}
To the best of the authors' knowledge, a  communication-theoretical analysis of a point-to-point IRS-assisted FSO link based on the Huygens-Fresnel principle was first conducted in \cite{WCNC}, which is the conference version of this paper. In contrast to \cite{WCNC}, in this paper, we partition the IRS into tiles and consider not only LP profiles but also QP profiles.  Moreover,  multi-link FSO systems, IRS sharing protocols, and the associated analysis were not investigated in \cite{WCNC}.

The remainder of this paper is organized as follows. The system and channel models are presented in Section \ref{Sec_System}. The  point-to-point IRS-assisted  FSO channel gain is derived based on the Huygens-Fresnel principle in Section \ref{Sec_Analysis}. Then, in Section \ref{Sec_Multilink}, we propose three   IRS sharing protocols for  multi-link FSO systems. The BER and outage probability of IRS-assisted multi-link FSO systems is analyzed in Section \ref{Sec_Perf}. Simulation results are presented in Section \ref{Sec_Sim}, and conclusions are drawn in Section \ref{Sec_concl}.

\textit{Notations:} Boldface lower-case and upper-case letters denote vectors and matrices, respectively.  Superscript $(\cdot)^T$ and $\mathbb{E}\{\cdot\}$
denote the transpose and expectation operators, respectively. $x\sim \mathcal{N}(\mu,\sigma^2)$ represents a  Gaussian random variable with mean $\mu$ and variance $\sigma^2$. $\mathbf{I}_n$ is the $n\times n$ identity matrix, $j$ denotes the imaginary unit, and $(\cdot)^*$ and $\mathcal{R}\{\cdot\}$ represent the complex conjugate and real part of a complex number, respectively. Moreover, $\text{erf}(\cdot)$ and $\text{erfi}(\cdot)$ are the error function and the imaginary error function, respectively. Furthermore, rotation matrices
$\mathbf{R}_y({\phi})=
\left(\begin{smallmatrix}
\cos(\phi) & 0 & -\sin(\phi)\\
0 & 1 & 0\\
\sin(\phi) & 0 & \cos(\phi)
\end{smallmatrix}\right)$
and
 $
\mathbf{R}_z({\phi})=
\left(\begin{smallmatrix}
\cos(\phi) & \sin(\phi) &0\\
-\sin(\phi) & 	\cos(\phi) & 0\\
0 & 0 & 1
\end{smallmatrix}\right)$   denote the counter-clockwise rotation by angle $\phi$  around the $y$- and $z$-axes, respectively.
  
\section{System and Channel Model}\label{Sec_System}
We consider  $N$ FSO transmitter-receiver pairs connected via a single optical IRS  where each transmitter is equipped with a LS and each  receiver is equipped with a PD and a lens, see Fig.~\ref{Fig:MA}. 
\subsection{System Model}
\begin{figure}
	\centering
	\includegraphics[width=0.55\textwidth]{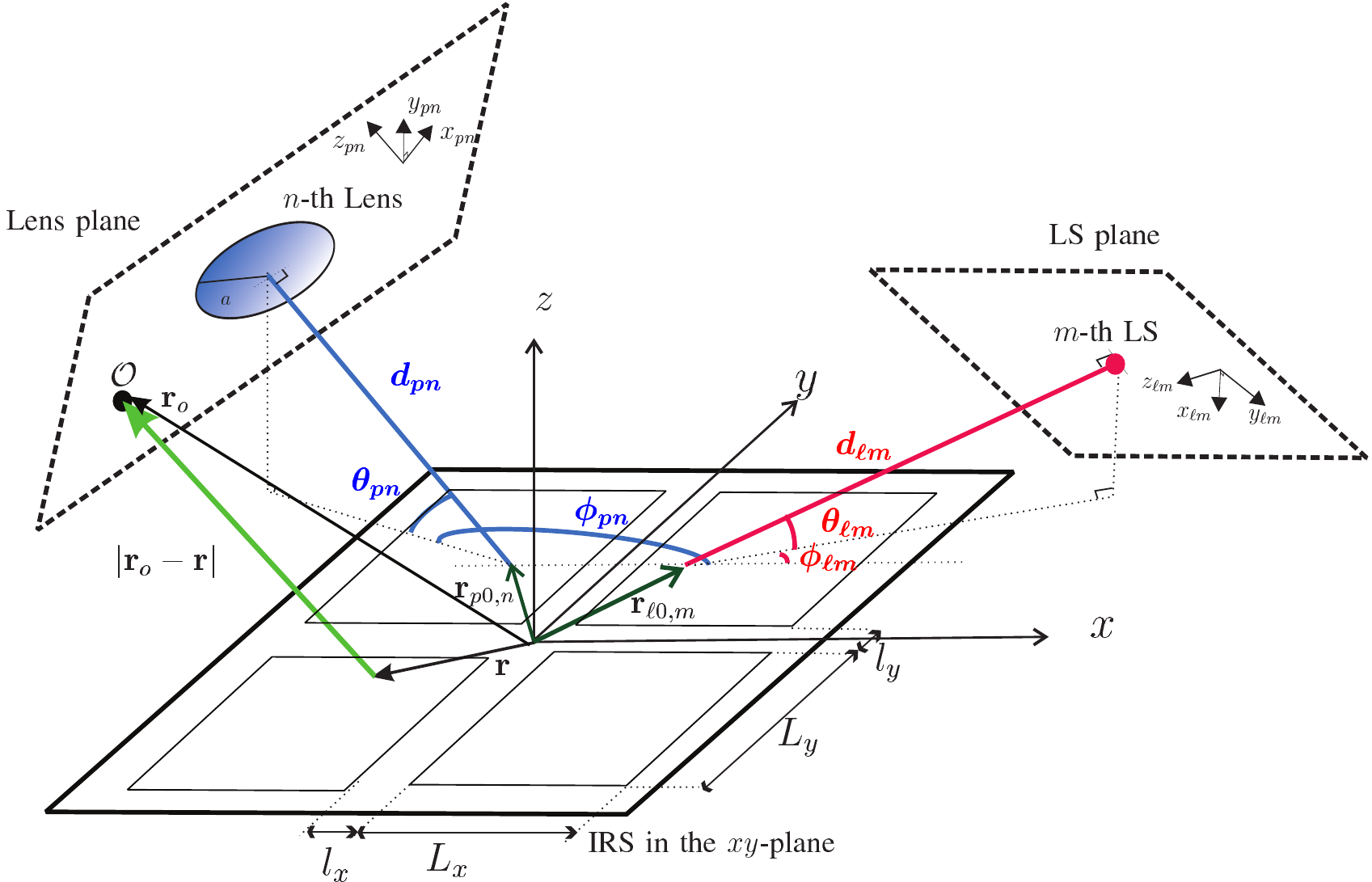}
	\caption{Schematic model of the IRS-assisted FSO system between the $m$-th LS and the $n$-th lens.}
	\label{Fig:System}\vspace*{-6mm}
\end{figure}

The IRS is  installed on a building wall which we define  as the $xy$-plane  and the center of the IRS is  at the origin, see Fig.~\ref{Fig:System}. The size of the IRS is $L_{x,\text{tot}}\times L_{y,\text{tot}}$ and it consists of  $Q=Q_xQ_y$ tiles, each comprising a large number of subwavelength unit cells, where $Q_x$ and $Q_y$ are the numbers of tiles in  $x$- and $y$-direction, respectively.  Each tile is centered  at  $\mathbf{r}_q=\left(x_q, y_q, 0\right)$ and has  length $L_x$ with tile spacing $l_x$ in  $x$-direction and width $L_y$ with tile spacing $l_y$ in  $y$-direction, see Fig.~\ref{Fig:System}. Given that  the size of a tile is much larger than the optical wavelength, i.e., $L_x, L_y \gg \lambda$, each tile  can be modeled as a continuous surface with a continuous phase shift profile centered at $\mathbf{r}_q^t$ and  denoted by $\Phi_{q}({\mathbf{r},\mathbf{r}_q^t})$ \cite{Vahid_IRS}, where $\mathbf{r}=(x,y,0)$ denotes a point in the $xy$-plane. Thus, the IRS length in  $x$- and $y$-direction is given by $L_{i,\,\text{tot}}=Q_iL_i+(Q_i-1)l_i$, $i\in\{x,y\}$. To employ the IRS for multi-link FSO transmission, we propose  different IRS sharing protocols  in Section \ref{Sec_IRSdesigns}.  As  shown in Fig.~\ref{Fig:System}, the $m$-th LS  is located at distance $d_{\ell m}, \, m\in \mathcal{N}$, where $\mathcal{N}=\{1, \dots, N\}$,  from the center of its beam footprint  on the IRS along the beam axis and its direction is denoted by $\mathbf{\Psi}_{\ell m}=\left(\theta_{\ell m},\phi_{\ell m}\right)$, where $\theta_{\ell m}$ is the angle between the $xy$-plane and the beam axis, and $\phi_{\ell m}$ is the angle between the projection of the beam axis on the $xy$-plane and the  $x$-axis. For simplicity, we assume all LSs are installed at the same height as the IRS and thus, $\phi_{\ell m}=0,\, \forall m \in\mathcal{N}$, see Fig.~\ref{Fig:MA}.  Moreover,   the $m$-th laser beam footprint  on the IRS plane is centered at point $\mathbf{r}_{\ell0, m}=\left(x_{\ell0,m}, y_{\ell 0,m},0\right)$. Furthermore, we assume the $n$-th PD is equipped with a circular lens of radius $a$ with  distance $d_{pn}, \, n\in\mathcal{N}$, from the IRS  along the normal vector of $n$-th the lens plane which has direction $\mathbf{\Psi}_{pn}=\left(\theta_{pn}, \phi_{pn}\right)$, where $\theta_{pn}$ is the angle between the $xy$-plane and the normal vector, and $\phi_{pn}$ is the angle between the projection of the normal vector on the $xy$-plane and the $x$-axis. The normal vector of the $n$-th lens plane intersects the IRS plane at point $\mathbf{r}_{p0,n}=\left(x_{p0,n},y_{p0,n},0\right)$, which we refer to as the lens center on the IRS. The $n$-th lens receives a  beam reflected from the IRS  and   focuses the beam   reflected  by the IRS  onto its PD. 
\subsection{Signal Model}
Assuming an intensity modulation and direct detection (IM/DD) FSO system, the  signal intensity $y_n[t]$ received by the $n$-th PD in  time slot $t, t\in \mathcal{T}$, where $\mathcal{T}=\{1, \dots, T\}$,  can be modeled as follows
\begin{IEEEeqnarray}{rll}
	y_n[t]=h_{n,n}  s_n[t]+\sum_{m=1, m\neq n }^N h_{m,n}  s_m[t]+w_n[t], \quad n\in\mathcal{N}, t\in\mathcal{T},
	\label{system_eq}
\end{IEEEeqnarray}
where  $s_m[t]$ is  the on-off keying (OOK) modulated symbol transmitted by the $m$-th LS with  average transmit power $P_{\ell m}$, $h_{m,n}\in \mathcal{R}^+$ is the channel gain between the $m$-th LS and the $n$-th PD, and  $w_n[t]\sim \left(0, \sigma_w^2\right)$ is the additive white Gaussian noise (AWGN)  with zero mean and variance $\sigma_w^2$ impairing  the $n$-th PD.
\subsection{Channel Model}
In general, FSO channels are affected by  geometric and misalignment losses (GML),  atmospheric losses, and  atmospheric turbulence induced fading \cite{ICC}.   
Thus, the IRS-assisted FSO channel gain $h_{m,n}$ between the  $m$-th LS  and the $n$-th PD  can be modeled as follows
\begin{IEEEeqnarray}{rll}
	h_{m,n}=h_p^{m,n} h_{\text{irs}}^{m,n} h_a^{m,n},
	\label{ch}
\end{IEEEeqnarray}
where $h_a^{m,n}$ represents the random atmospheric turbulence induced fading,  $h_p^{m,n}=10^{-\frac{\kappa}{10}(d_{\ell m}+d_{pn})}$ is the atmospheric loss which depends on the attenuation coefficient, $\kappa$, and  $h_{\text{irs}}^{m,n}$ characterizes the deterministic GML. Here, we assume that the $h_a^{m,n}, \forall m,n\in\mathcal{N}$, are independent and non-identically distributed Gamma-Gamma variables, i.e., $h_a^{m,n}\sim \mathcal{GG}(\alpha_{m,n}, \beta_{m,n})$, with  small and large scale turbulence parameters $\alpha_{m,n}$ and $\beta_{m,n}$ \cite{GG_Murat}. Moreover, $h_\text{irs}^{m,n}$ denotes the fraction of the power of the $m$-th LS  that is  reflected by the IRS and collected by the $n$-th PD, i.e.,
\begin{IEEEeqnarray}{rll}
	h_\text{irs}^{m,n}={1 \over P_{\ell m}}\iint\limits_{\mathcal{A}_{pn}} I_\text{irs}^{m,n}\!\!\left(\mathbf{r}_{pn}\right)\, \mathrm{d}\mathcal{A}_{pn},
	\label{gain}
\end{IEEEeqnarray}
where  $\mathcal{A}_{pn}$ denotes the area of the lens of the $n$-th PD,  $I_\text{irs}^{m,n}\!\!\left(\mathbf{r}_{pn}\right)$ is  the power intensity of the  beam emitted by the $m$-th LS and reflected by the IRS   in  the  plane of the $n$-th lens, and $\mathbf{r}_{pn}=(x_{pn}, y_{pn}, z_{pn})$ denotes    a point on the lens plane.  The origin of the $x_{pn}y_{pn}z_{pn}$-coordinate system  is  the center of the $n$-th lens and the $z_{pn}$-axis  is parallel to  the normal vector of the lens plane, see Fig.~\ref{Fig:System}. We assume that the $y_{pn}$-axis is parallel to the intersection line of the lens plane and the IRS plane and the $x_{pn}$-axis is perpendicular to the $y_{pn}$- and $z_{pn}$-axes.

Assuming that the waist of the Gaussian laser beam, $w_{0m}$, is larger than the wavelength, $\lambda$, the paraxial approximation is valid and the power intensity of the  reflected beam is given as follows  \cite{saleh}
\begin{IEEEeqnarray}{rll}
	I_\text{irs}^{m,n}\left(\mathbf{r}_{pn}\right)=\frac{1}{2\eta}\lvert E_\text{irs}^{m,n}\left(\mathbf{r}_{pn}\right) \rvert^2,
	\label{irradiance}
\end{IEEEeqnarray}
where $\eta$ is the free-space impedance and $E_\text{irs}^{m,n}(\mathbf{r}_{pn})$ is the  electric field emitted by the $m$-th LS, reflected by the IRS, and observed by the $n$-th lens. Thus, we have 
\begin{IEEEeqnarray}{rll}
	E_\text{irs}^{m,n}\left(\mathbf{r}_{pn}\right)=\sum_{q=1}^Q E_q^{m,n}\left(\mathbf{r}_{pn}\right),
	\label{E_r}
\end{IEEEeqnarray}
 where $E_q^{m,n}\left(\mathbf{r}_{pn}\right)$ is the part of the  electric field of the $m$-th LS which is reflected  by tile $q$, see Section \ref{Sec_Analysis}. The electric field of the Gaussian laser beam  emitted by the  $m$-th LS   is given by \cite{Goodmanbook}
 \begin{IEEEeqnarray}{rll}
 	E_{\ell m}\!\left(\mathbf{r}_{\ell m}\right)=& \frac{E_{0m}w_{0m}}{{w}(z_{\ell m})} \exp\left(-{x_{\ell m}^2+y_{\ell m}^2\over w^2(z_{\ell m})}-j\psi_{\ell m}\right)\quad\text{with phase} \nonumber\\
 	\psi_{\ell m}&=k\left(z_{\ell m}+{x_{\ell m}^2+y_{\ell m}^2\over 2R(z_{\ell m})}\right)-\tan^{-1}\left(\frac{z_{\ell m}}{z_{0m}}\right),\quad
 	\label{Gauss}
 \end{IEEEeqnarray}
 where $\mathbf{r}_{\ell m}=(x_{\ell  m}, y_{\ell  m}, z_{\ell m})$ is a point in the  coordinate system which has its origin at the $m$-th LS.  The $z_{\ell m}$-axis of this coordinate system  is along the beam axis, its $y_{\ell m}$-axis is parallel to the intersection line of the LS plane and the IRS plane, and its $x_{\ell m}$-axis  is orthogonal to the $y_{\ell m}$- and $z_{\ell m}$-axes. Here, $E_{0m}$ is the  electric field at the origin of the $x_{\ell m}y_{\ell m}z_{\ell m}$-coordinate system, $k=\frac{2\pi}{\lambda}$ is the wave number, $\lambda$ denotes the wavelength, $w(z_{\ell m})=w_{0m}\left[{1+\left(\frac{z_{\ell m}}{z_{0m}}\right)^2}\right]^{1/2}$ is the beam width at distance $z_{\ell m}$, $R(z_{\ell m})=z_{\ell m}\left[1+\left(\frac{z_{0m}}{z_{\ell m}}\right)^2\right]$ is the radius of the curvature of the beam's wavefront, and $z_{0m}=\frac{\pi w_{0m}^2}{\lambda}$ is the Rayleigh range. Here, the total  power emitted by the $m$-th LS is given by $P_{\ell m}=\frac{\pi}{4\eta}|E_{0m}|^2w^2_{0m}$.
 
\section{Point-to-Point Optical IRS Channel Model}\label{Sec_Analysis}

In this section, we focus on a single  LS-PD pair to model the impact of the IRS on the end-to-end channel. To simplify the notation, in this section, we drop the index of the LS $(m)$ and the index of the PD $(n)$ in all variables. In the following, using the  electric field of the LS in (\ref{Gauss}), $E_\ell\left(\mathbf{r}_\ell\right)$, first, we determine the  electric field incident on the IRS, $E_{\text{in}}(\mathbf{r})$. Then, we specify the electric field  reflected from the $q$-th tile, $E_q\left(\mathbf{r}_{p}\right)$. Finally,  we determine the total electric field reflected from the IRS and the point-to-point channel gain between the LS and the PD, $h_{\text{irs}}$.

\subsection{Incident Beam}
First,  in the following lemma, we determine the   electric field incident on the IRS plane.
\begin{lem}\label{Lemma1}
 	Assuming that $d_\ell\gg L_{x}, L_{y}$, the  electric field  emitted by the LS incident on  the IRS plane, denoted by $E_{\text{in}}({\mathbf{r}})$,  is given by 
 	\begin{IEEEeqnarray}{rll}
 	E_{\text{in}}({\mathbf{r}})&= \frac{E_0 w_0\zeta_\text{in}}{{w}(\hat{d}_\ell)}\exp\left(-{\hat{x}^2\over w_x^2(\hat{d}_\ell)}-{\hat{y}^2\over w_y^2(\hat{d}_\ell)}-j\psi_\text{in}({\mathbf{r}})\right)\quad\text{with phase}\\
 			\psi_{\text{in}}({\mathbf{r}})&=k\left(\hat{d}_\ell-{x}\cos(\theta_\ell)+{\hat{x}^2\over 2R_x(\hat{d}_\ell)}+{\hat{y}^2\over 2R_y(\hat{d}_\ell)}\right)-\tan^{-1}\left(\frac{\hat{d}_\ell}{z_0}\right),\quad
\label{lem1}
 \end{IEEEeqnarray} 
where $\zeta_\text{in}=\sqrt{\lvert\sin(\theta_\ell)\rvert}$, $\hat{d}_\ell=d_\ell+{x}_{\ell 0}\cos(\theta_\ell)$, $\hat{\mathbf{r}}=\left(\hat{x},\hat{y},0\right)={\mathbf{r}}-{\mathbf{r}}_{\ell 0}$,  $w_x(\hat{d}_\ell)=\frac{w(\hat{d}_\ell)}{\sin(\theta_\ell)}$, 	$w_y(\hat{d}_\ell)=w(\hat{d}_\ell)$, $R_x(\hat{d}_\ell)=\frac{R(\hat{d}_\ell)}{\sin^2(\theta_\ell)}$, and $R_y(\hat{d}_\ell)=R(\hat{d}_\ell)$.
\end{lem}
\begin{IEEEproof}
The proof is given in Appendix \ref{App0}.
\end{IEEEproof}

Eq.~(\ref{lem1}) describes an elliptical Gaussian beam on the IRS with beam widths  $w_x$ and $w_y$  along the $x$- and $y$-axes, respectively. 

\subsection{Huygens-Fresnel Principle}
 To determine the impact of the IRS on the incident beam, we use scalar field theory \cite{Goodmanbook} and neglect the vectorial nature of the electromagnetic field. The scalar field  approach yields  accurate results if the following conditions are met: 1) the diffracting surface must be large compared to the wavelength, 2) the electromagnetic fields must not be observed very close to the surface, i.e., $d_p\gg \lambda$ \cite{Goodmanbook}. Given the size of the IRS and the use of   FSO systems for long-distance communications, these conditions are met in practice. Thus, we can apply the Huygens-Fresnel principle \cite{Goodmanbook},  a scalar field analysis method,  for deriving  the  beam reflected by the IRS. This principle states that  every point on the wavefront of the beam can be considered as a secondary source emitting a  spherical wave and, at any position, the new wavefront  is determined by the sum of these  secondary waves \cite{Goodmanbook}.  Given this principle, the complex amplitude of the electric field  reflected by the $q$-th tile, denoted by $E_{q}(\mathbf{r}_{o})$, at an arbitrary observation point $\mathcal{O}$ located at $\mathbf{r}_o=(x_o, y_o, z_o)$, see Fig.~\ref{Fig:System},   is given by \cite{Goodmanbook}
\begin{IEEEeqnarray}{rll}
	E_{q}(\mathbf{r}_{o})&=\frac{1}{j\lambda}\iint_{({x},{y})\in\Sigma_q} E_{\text{in}}({\mathbf{r}}) S\left({\mathbf{r}},\mathbf{r}_o\right)  T_q({\mathbf{r}}) \mathrm{d}{x}\mathrm{d}{y},\,\quad
	\label{Huygens-Fresnel}
\end{IEEEeqnarray}
where   $S\left({\mathbf{r}},\mathbf{r}_o\right)={\exp(jk|\mathbf{r}_{o}-{\mathbf{r}}|)\over |\mathbf{r}_{o}-{\mathbf{r}}|}$ represents a spherical wave, see \cite[Eq.~(3-49)]{Goodmanbook} , $T_q(\mathbf{r})=\zeta_{q} e^{-j\Phi_{q}({\mathbf{r},\mathbf{r}_q^t})}$ is  the tile response, and $\Sigma_q$  is   the tile area. Here,   $\zeta_{q}$ denotes the efficiency factor ($0\leq\zeta_q\leq 1$), which accounts for the portion of incident power propagated towards the  lens  and   $\Phi_{q}(\mathbf{r},\mathbf{r}_q^t)$ is the phase shift profile of the $q$-th tile centered at $\mathbf{r}_q^t$. In (\ref{Huygens-Fresnel}), the total surface of the tile is divided into infinitesimally small areas $\mathrm{d}{x}\mathrm{d}{y}$, and the light wave scattered by each  area is modeled as a secondary source emitting a spherical wave, modeled by ${\exp(jk|\mathbf{r}_{o}-{\mathbf{r}}|)\over |\mathbf{r}_{o}-\mathbf{r}|}$. The complex amplitudes of the secondary sources are proportional to the incident electric field, $E_{\text{in}}({\mathbf{r}})$, and an additional phase shift term, $e^{j\Phi_{q}({\mathbf{r},\mathbf{r}_q^t})}$, is introduced by the tile.  The phases of the spherical sources, $k|\mathbf{r}_o-{\mathbf{r}}|$, play an important role in our analysis, and to find a closed-form solution for the integral in (\ref{Huygens-Fresnel}), we approximate $|\mathbf{r}_o-{\mathbf{r}}|$ in the following.  
\subsection{Intermediate-Field vs. Far-Field}\label{Sec_IntvsFar}
First, using $|\mathbf{r}_o-\mathbf{r}|=\left[({x}-x_o)^2+({y}-y_o)^2+z_o^2\right]^{1/2}$, we establish
\begin{IEEEeqnarray}{rll}	
	&\frac{|\mathbf{r}_o-{\mathbf{r}}|^2}{|{\mathbf{r}}_o|^2}=1+\frac{x^2+y^2}{|{\mathbf{r}}_o|^2}-2\frac{xx_o+yy_o}{|{\mathbf{r}}_o|^2}.\qquad
\end{IEEEeqnarray}
Applying the Taylor series expansion \cite{integral} with $(1+\varkappa)^{1/2}=1+\frac{1}{2} \varkappa-\frac{1}{8} \varkappa^2+ \cdots$, we obtain
\begin{IEEEeqnarray}{rll}	
{|\mathbf{r}_o-{\mathbf{r}}|}&=\underbrace{|{\mathbf{r}}_o|-{\frac{xx_o+yy_o}{|{\mathbf{r}}_o|}}}_{=\text{t}_1}+\underbrace{\frac{x^2+y^2}{2|{\mathbf{r}}_o|}-\frac{x^2x_o^2+y^2y_o^2}{2|{\mathbf{r}}_o|^3}}_{=\text{t}_2}\nonumber\\
&-\underbrace{\frac{(x^2+y^2)^2}{8|{\mathbf{r}}_o|^3}+\frac{(x^2+y^2)(xx_o+yy_o)}{2|{\mathbf{r}}_o|^3}-\frac{xyx_oy_o}{|{\mathbf{r}}_o|^3}}_{=\text{t}_3}+\cdots
	.\qquad
	\label{bionom}
\end{IEEEeqnarray}
For the commonly used far-field approximation, it is assumed that the secondary waves reflected by the tile surface  experience  only a linear phase shift w.r.t. each other \cite{Hecht}. This approximation is equivalent  to  assuming a linear phase shift for the phase of the secondary sources w.r.t.  the $x$- and $y$-directions. In other words, only $\text{t}_1$ in (\ref{bionom}) is taken into account, and     $\text{t}_2$ and all higher orders terms are  neglected. For the far-field approximation to hold,  the impact of   $\text{t}_2$ in the argument of the  exponential term,  $k|\mathbf{r}_{o}-{\mathbf{r}}|$, should be much smaller than one period of the complex exponential, and thus, 
  \begin{IEEEeqnarray}{rll}	
k\frac{x^2+y^2}{2|\mathbf{r}_{o}|}\ll 2\pi.
\label{phass}
\end{IEEEeqnarray}
The range of the  relevant values for $x$ and $y$ in  (\ref{Huygens-Fresnel}) is bounded by the beam widths of the incident electric field $2w_x$ and $2w_y$ (where the power of the incident beam drops by $\frac{1}{e^4}$ compared to the peak value)    and the size of the tile $L_{x}$ and $L_{y}$, i.e., $x_e=\min\left(\frac{L_{x}}{2}, w_x\right)\geq |x|$ and $y_e=\min\left(\frac{L_{y}}{2}, w_y\right)\geq |y|$. Assuming that the lens radius is smaller than the IRS-lens distance, i.e., $a\ll d_p$, for any observation point $\mathcal{O}$  on the lens area, we can substitute  $|\mathbf{r}_{o}|\approx d_p$. Thus,   we substitute $x_e$ and $y_e$ for $x$ and $y$ in (\ref{phass}), respectively,  and  define the minimum far-field distance $d_f$ as follows
 \begin{IEEEeqnarray}{rll}	
d_f=\frac{x_e^2+y_e^2}{2\lambda},\qquad
\label{d_f}
\end{IEEEeqnarray}
 such that for  distances $d_p\gg d_f$, the approximation of (\ref{bionom}) in (\ref{Huygens-Fresnel}) by   term $\text{t}_1$ only is appropriate.  However, depending  on the values of $x_e$ and $y_e$, and the IRS-lens  distance, $d_p$, this condition might not hold in practice. For example, consider a typical IRS with tile size $L_{x}=L_{y}=50$ cm and a LS located at $ d_\ell=1000$ m and $\mathbf{\Psi}_\ell=\left( \frac{\pi}{8}, 0\right)$. Then, a laser beam with wavelength  $\lambda=1550$ nm and $w_0=2.5$ mm has widths  $w_x=0.52$ m and $w_y=0.19$ m on the IRS and the minimum far-field distance according to (\ref{d_f}) is $d_f=32.7$ km, which exceeds the typical link distance of FSO systems. Thus, in order to  obtain a model that is also valid for   shorter distances, we  have to consider both the linear term $\text{t}_1$ and the quadratic term $\text{t}_2$ in (\ref{bionom}). Using a similar method as in (\ref{phass}),  we define intermediate distances as the range where the largest term in $\text{t}_3$ is smaller  than  a wavelength, i.e.,
\begin{IEEEeqnarray}{rll}	
k\frac{(x^2+y^2)(xx_o+yy_o)}{2|{\mathbf{r}}_o|^3}\ll 2\pi.
\label{phass2}
\end{IEEEeqnarray}
Substituting $x_e$ and $y_e$ for $x$ and $y$ and setting $x_o=y_o={|{\mathbf{r}}_o|\over 2}\approx{d_p\over 2}$, we define the minimum intermediate distances $d_n$ as follows
\begin{IEEEeqnarray}{rll}	
d_n=\left[\frac{(x_e^2+y_e^2)(x_e+y_e)}{4\lambda}\right]^{1/2}.
\label{d_n}
\end{IEEEeqnarray}
 For the previous example,  we obtain $d_n=85.6$ m which is much smaller than the practical range of link distances of FSO systems. Thus, for practical link distances $d_p\gg d_n$, the approximation of (\ref{bionom}) in (\ref{Huygens-Fresnel}) by  terms $\text{t}_1$ and $\text{t}_2$  is appropriate.  
 \subsection{Received Electric Field  from a Tile}
 As mentioned previously, the Gaussian beam incident on the IRS in (\ref{lem1}) has a nonlinear phase profile. In order to redirect the beam  in a desired direction, the IRS must compensate the phase of the incident beam and apply an additional  phase shift to redirect the beam. In the following theorem,  we assume a QP profile across the IRS as an approximation of more general nonlinear phase shift profiles and  obtain  a closed-form solution for the reflected electric field  in (\ref{Huygens-Fresnel}).
\begin{theo}\label{Theorem1}
Assume a QP profile  centered at point $\mathbf{r}_q^t$, i.e.,  $\Phi_q^{\text{quad}}({\mathbf{r}},\mathbf{r}_q^t)=k\big(\Phi_{q,0}+\Phi_{q,x} ({x}-x_q^t)$$+\Phi_{q,y} (y-y_q^t)+\Phi_{q,x^2} ({x}-x_q^t)^2+\Phi_{q,y^2} ({y}-y_q^t)^2\big)$, where $\Phi_{q,0}$, $\Phi_{q,x}$, $\Phi_{q,x^2}$, $\Phi_{q,y}$, and $\Phi_{q,y^2}$ are  constants. Then, the  electric field emitted by the LS  at position $\left(d_\ell,\mathbf{\Psi}_\ell\right)$, reflected by the tile   centered at  $\mathbf{r}_q=\left(x_q,y_q,0\right)$, and received  at the lens located at $\left(d_p, \mathbf{\Psi}_p\right)$   for any intermediate distances $d_\ell\gg L_{x},L_{y}$ and $d_p \gg a, x_{\ell0}, y_{\ell0},d_n$ is given by  
\begin{IEEEeqnarray}{rll}				
	&\hspace*{-4mm}E_{q}(\mathbf{r}_p)={CC_q} {\frac{\pi}{4\sqrt{b_x b_y}}}\, e^{-\frac{k^2}{4b_x}X^2-\frac{k^2}{4b_y}Y^2}\nonumber\\
	&\hspace*{-2mm}\times\Bigg[\text{erf}\left({\sqrt{b_x}}\left(x_q+\frac{L_x}{2}\right)+\frac{jk}{2\sqrt{b_x}} X\right)-\text{erf}\left({\sqrt{b_x}}\left(x_q-\frac{L_x}{2}\right)+\frac{jk}{2\sqrt{b_x}}X\right)\Bigg]\nonumber\\
	&\hspace*{-2mm}\times\Bigg[\text{erf}\left({\sqrt{b_y}}\left(y_q+\frac{L_y}{2}\right)+\frac{jk}{2\sqrt{b_y}}Y\right)-\text{erf}\left({\sqrt{b_y}}\left(y_q-\frac{L_y}{2}\right)+\frac{jk}{2\sqrt{b_y}}Y\right)\Bigg],\quad
	\label{theo1}
\end{IEEEeqnarray}
where $X=A_0+c_1x_p+c_2y_p+\Phi_{q,x}-2x_q^t\Phi_{q,x^2}$, $A_0={2j\nu{x}_{\ell0}\over k}\sin^2(\theta_\ell)+\frac{{x}_{p0}}{d_p}+\varphi_x$,
 $Y=B_0+c_3x_p+c_4y_p+\Phi_{q,y}-2y_q^t\Phi_{q,y^2}$, $B_0={2j\nu{y}_{\ell0}\over k}+\frac{{y}_{p0}}{d_p}+\varphi_y$,  $c_1 = \frac{1}{d_p}\cos(\phi_p)\sin(\theta_p)$, $c_2=-\frac{1}{d_p}\sin(\phi_p)$, $c_3=\frac{1}{d_p}\sin(\phi_p)\sin(\theta_p)$, $c_4=\frac{1}{d_p}\cos(\phi_p)$, $c_5=\frac{1}{d_p}\cos(\phi_p)\cos(\theta_p)$, $c_6=\frac{1}{d_p}\sin(\phi_p)\cos(\theta_p)$,  $\varphi_x=-\cos(\theta_\ell)-\cos(\theta_p)\cos(\phi_p)$,  $\varphi_y=-\cos(\theta_p)\sin(\phi_p)$, $b_x=\nu\sin^2(\theta_\ell)-\frac{jk}{2d_p}\left(1-c_5^2d_p^2-2c_5x_{p0}\right)-jk\Phi_{q,x^2}$,
 $b_y=\nu-\frac{jk}{2d_p}\left(1-c_6^2d_p^2-2c_6y_{p0}\right)-jk\Phi_{q,y^2}$, $\delta_{q}=-\Phi_{q,x^2}(x_q^t)^2-\Phi_{q,y^2}(y_q^t)^2+\Phi_{q,x}x_q^t+\Phi_{q,y} y_q^t-\Phi_{q,0}$, $C_q=\zeta_q e^{jk\delta_{q}}$, $\nu={1\over w^2(\hat{d}_\ell)}+{jk\over 2R(\hat{d}_\ell)}$, $C=\frac{E_ow_0\zeta_\text{in}}{j\lambda{w}(\hat{d}_\ell)d_p}e^{-jk(\hat{d}_\ell-d_p)+j\tan^{-1}\left(\frac{\hat{d}_\ell}{z_0}\right)-\nu\sin^2(\theta_\ell) {x}_{\ell0}^2-\nu {y}_{\ell0}^2}$.
\end{theo}
\begin{IEEEproof}
The proof is given in Appendix \ref{App1}.
\end{IEEEproof}
Eq.~(\ref{theo1})  explicitly shows the impact of the positioning of the LS and the lens w.r.t. the IRS,  the size of the tile, and  the phase shift configuration across the tile on the electric field reflected by the tile. The above theorem is valid  for both far-field and intermediate distances.  The electric field for LP profiles is included in  the result in (\ref{theo1}) as a special case for $\Phi_{{q,x^2}}=\Phi_{{q,y^2}}=0$, see Section \ref{Sec_Tile_Config} for more details on the tile phase shift configuration.

In the following corollary, as a special case of Theorem \ref{Theorem1}, we consider the conventional mirror. A conventional mirror introduces  no additional phase shifts, i.e., $\Phi_q({\mathbf{r}},\mathbf{r}_q^t)=0$, and the incident angle and the reflection angle follow Snell's law, i.e., $\theta_\ell=\theta_p$. 
\begin{corol}[Reflection by Conventional Mirror]\label{col1}
Assume a large conventional mirror  or equivalently a large tile of an IRS, i.e., ${L_x}, L_y\gg 2w(\hat{d_\ell})$, such that  the entire received beam is reflected.  Then, assuming a  far-field scenario, $d_p\gg d_f$, $\phi_p=\pi$, and $x_{\ell0}=y_{\ell0}=0$,  (\ref{theo1}) simplifies to	
	\begin{IEEEeqnarray}{rll}
		E_{q}(\mathbf{r}_p)&=\frac{E_0 w_0\zeta_{t}}{w_\text{cir}} \exp\left(-\frac{x_p^2+y_p^2}{w_\text{cir}^2}-j\psi_\text{cir}\right) \text{with phase}\nonumber\\
		& \psi_\text{cir}=k\left(\hat{d}_\ell-d_p-\frac{x_p^2}{2R_\text{cir}}-\frac{y_p^2}{2R_\text{cir}}\right)+\frac{\pi}{2}-\tan^{-1}\left(\frac{\hat{d}_\ell}{z_0}\right),
		\label{GEo-Gauss}
	\end{IEEEeqnarray}
where $\zeta_{t}=\frac{\zeta_\text{in}\zeta_q}{|\sin(\theta_\ell)|}$, and  $w_\text{cir}=\frac{2w(\hat{d}_\ell) |\nu|d_p}{k}$ and  $R_\text{cir}=\frac{4d_p^2|\nu|^2 R(\hat{d}_\ell)}{k^2}$ are the equivalent circular beamwidth and radius of curvature, respectively. 
\end{corol}
\begin{IEEEproof}
Considering ${L_x}, L_y\gg 2w(\hat{d}_\ell)$, we can substitute the $\text{erf}(\cdot)$-terms in (\ref{theo1}) by 4. Assuming  $d_p\gg d_f$, we obtain $b_x=\nu\sin^2(\theta_\ell)$ and $b_y=\nu$. Substituting $\theta_\ell=\theta_p$, $\phi_p=\pi$, and $\Phi_q({\mathbf{r}},\mathbf{r}_q^t)=0$ leads to (\ref{GEo-Gauss}) and this completes the proof.
\end{IEEEproof}
Eq.~(\ref{GEo-Gauss}) corresponds to a circular Gaussian beam, and reveals that, in the considered regime, the reflected beam is identical to what  is expected from geometric optics for a conventional mirror, see \cite{Marzieh_IRS,Marzieh_IRS_jou}.  
\begin{corol}[Reflection by Anomalous Mirror]\label{col2}
	We assume an anomalous mirror which can  impose an additional linear phase shift, i.e., $\Phi_{q,0}=0$, $\Phi_{q,x}=\cos(\theta_\ell)\cos(\phi_\ell)+\cos(\theta_p)\cos(\phi_p)$,  $\Phi_{q,y}=\cos(\theta_\ell)\sin(\phi_\ell)+\cos(\theta_p)\cos(\phi_p)$, $\Phi_{q,x^2}=\Phi_{q,y^2}=0$.  Then, for the far-field scenario, i.e., $d_p\gg d_f$,   ${L_x}, L_y\gg 2w(\hat{d_\ell})$, and  $x_{\ell 0}=y_{\ell0}=0$,  (\ref{theo1}) simplifies to
\begin{IEEEeqnarray}{rll}
	E_{q}(\mathbf{r}_p)&= \frac{E_0 w_0\zeta_{t}}{\sqrt{w_{x,\text{elp}}w_{y,\text{elp}}}}\sqrt{\frac{\sin(\theta_\ell)}{\sin(\theta_p)}}\exp\left(-\frac{x_p^2}{w_{x,\text{elp}}^2}-\frac{y_p^2}{w^2_{y,\text{elp}}}-j\psi_\text{elp}\right) \text{with phase}\nonumber\\
	&\psi_\text{elp}=k\left(\hat{d}_\ell-d_p-\frac{x_p^2}{2R_{x,\text{elp}}}-\frac{y_p^2}{2R_{y,\text{elp}}}\right)+\frac{\pi}{2}-\tan^{-1}\left(\frac{\hat{d}_\ell}{z_0}\right),
		\label{ellips_Gaus}
	\end{IEEEeqnarray}
where $w_{x,\text{elp}}=\frac{w_{y,\text{elp}}|\sin(\theta_\ell)|}{|\sin(\theta_p)|}$, $w_{y,\text{elp}}=\frac{2|\nu|d_pw(\hat{d}_\ell)}{k}$, $R_{x,\text{elp}}=\frac{R_{y,\text{elp}}\sin^2(\theta_\ell)}{\sin^2(\theta_p)}$, and $R_{y,\text{elp}}=\frac{4d_p^2|\nu|^2 R(\hat{d}_\ell)}{k^2}$.
\end{corol}
\begin{IEEEproof}
Substituting $\Phi_{q,x}$ and  $\Phi_{q,y}$ in (\ref{theo1}) and  following similar steps as in the proof of Corollary \ref{col1} but for the general case  $\theta_\ell\neq\theta_p$ leads to (\ref{ellips_Gaus}). This completes the proof. 
\end{IEEEproof}
Eq.~(\ref{ellips_Gaus}) describes an elliptical Gaussian beam. This result is  in agreement with the result from geometric optics  for anomalous reflection \cite{Marzieh_IRS_jou}.

Depending on the tile size,   distances $d_p\gg d_f$ might not be in the practical range of  FSO systems, see Section \ref{Sec_Sim}. Therefore,  Corollaries \ref{col1} and \ref{col2},  which are valid if the far-field approximation holds,  do not always provide accurate results.  Thus, in general, Theorem \ref{Theorem1}  is required to determine the electric field for  practical applications. Moreover, by choosing appropriate values for  $\Phi_{q,i}$, $i\in\{0,x,y,x^2,y^2\}$ for the QP profile across the tile in (\ref{theo1}), we can redirect the reflected electric field in a desired direction and reduce the divergence of the beam along the propagation path, see Section \ref{Sec_Tile_Config}.
 
\subsection{Point-to-Point GML}
In the previous subsection, we have determined the  electric field reflected from a tile centered at  $\mathbf{r}_q$ with phase shift profile $\Phi_q({\mathbf{r}},\mathbf{r}_q^t)$ across the tile. Now,  using (\ref{theo1}), (\ref{E_r}), and the power intensity  in (\ref{irradiance}), the  deterministic GML in  (\ref{gain}) can be rewritten as follows
\begin{IEEEeqnarray}{rll}
h_\text{irs}=\frac{1}{2\eta P_\ell} \iint_{\mathcal{A}_{p}} \Big\lvert\sum_{q=1}^Q E_q\left(\mathbf{r}_{p}\right)\Big\rvert^2 \mathrm{d}\mathcal{A}_{p}.
\label{Final-gain}
\end{IEEEeqnarray}
 
  In the following theorem, we simplify (\ref{Final-gain}) and determine  $h_\text{irs}$.
\begin{theo}[Out-of-Plane Reflection]\label{Theorem4}
	Assume $d_\ell\gg L_x,L_y$ and $d_p\gg a, x_{\ell0}, y_{\ell0}, d_n$. Then, the point-to-point deterministic GML, $h_{\text{irs}}$, including  geometric and deterministic misalignment losses, between  an LS at position $\left(d_\ell,\mathbf{\Psi}_\ell\right)$ illuminating an IRS in the $xy$-plane and  a receiver lens at position $\left(d_p,\mathbf{\Psi}_p\right)$ is given by
\begin{IEEEeqnarray}{rll}
			h^{\text{out}}_{\text{irs}}=&\frac{{C}_h\sqrt{\pi}}{2\sqrt{\rho_x}} \sum\limits_{q=1}^Q\!\sum\limits_{\varsigma=1}^Q C_q C_\varsigma^* C_{2,q} C_{2,\varsigma}^* \exp\!\left(\!-\frac{k^2}{4}\left(\frac{{A}_{q}^2}{b_x}+\frac{{A}_{\varsigma}^2}{{b}_x^*}+\frac{{B}_{q}^2}{b_y}+\frac{{B}_{\varsigma}^2}{{b}_y^*}\right)\!\right) \int\limits_{-\tilde{a}}^{\tilde{a}}   e^{-\rho_yy_p^2-\varrho_{y}y_p+\frac{(\rho_{xy}y_p+\varrho_{x})^2}{4\rho_x}}\nonumber\\ &\times\Big[\text{erf}\left(\frac{{\rho_x\sqrt{\pi}a}+\rho_{xy} y_p +\varrho_{x}}{2\sqrt{\rho_x}}\right)\!-\text{erf}\left(\frac{{-\rho_x\sqrt{\pi}a}+\rho_{xy} y_p +\varrho_{x}}{2\sqrt{\rho_x}}\right)\Big] \mathrm{d}y_p,\quad
			\label{theo4}
\end{IEEEeqnarray}
where $\tilde{a}=\frac{\sqrt{\pi}a}{2}$, $\tilde{b_i}={b_i{b}_i^*\over b_i+{b}_i^*},$ $i\in \{x, y\}$, $C_h\!=\!\frac{\pi\zeta^2_\text{in}}{8{|b_x||b_y|}\lambda^2d_p^2w^2(\hat{d}_\ell)} e^{-\frac{2x_{\ell0}^2\sin(\theta_\ell)}{w^2(\hat{d}_\ell)}-\frac{2y_{\ell0}^2}{w^2(\hat{d}_\ell)}}$, $\rho_x= \frac{k^2}{4}\left(\frac{c_1^2}{\tilde{b}_x}+\frac{c_3^2}{\tilde{b}_y}\right)$, $\rho_y=\frac{k^2}{4} \left(\frac{c_2^2}{\tilde{b}_x}+\frac{c_4^2}{\tilde{b}_y}\right)$, $\rho_{xy}=\frac{k^2}{2} \left(\frac{c_1c_2}{\tilde{b}_x}+\frac{c_3c_4}{\tilde{b}_y}\right)$, ${A}_i=A_0+\Phi_{i,x}-2x_i^t\Phi_{i,x^2}$,  ${B}_i=B_0+\Phi_{i,y}-2y_i^t\Phi_{i,y^2}$,
 $\varrho_{x}=\frac{k^2}{2}\left(c_1\left(\frac{{A}_{q}}{b_x}+\frac{{A}_{\varsigma}^*}{{b}_x^*}\right)+c_3\left(\frac{{B}_{q}}{b_y}+\frac{{B}_{\varsigma}^*}{{b}_y^*}\right)\right)$,  $\varrho_{y}=\frac{k^2}{2}\left(c_2\left(\frac{{A}_{q}}{b_x}+\frac{{A}_{\varsigma}^*}{{b}_x^*}\right)+c_4\left(\frac{{B}_{q}}{b_y}+\frac{{B}_{\varsigma}^*}{{b}_y^*}\right)\right)$,  
 $C_{2,i}=\left(\text{erf}\left(\sqrt{b_x}{L}_{x}-\varpi_{x,i}\right)-\text{erf}\left(-\varpi_{x,i} \right)\right)$$\left(\text{erf}\left(\sqrt{b_y}{L}_{y}-\varpi_{y,i}\right)-\text{erf}\left(-\varpi_{y,i}\right)\right)$, $\varpi_{x,i}={\sqrt{b_x}}\left(\frac{L_x}{2}-x_i\right)-\frac{jk}{2\sqrt{b_x}}\left(A_0+\left(c_1+c_2\right)\frac{a}{2}+\Phi_{i,x}-2x_i^t\Phi_{i,x^2}\right)$,
$\varpi_{y,i}\!=\!\!{\sqrt{b_y}}\left(\frac{L_y}{2}-y_i\right)-\frac{jk}{2\sqrt{b_y}}\big( B_0+\left(c_3+c_4\right)\frac{a}{2}+\Phi_{i,y}-2y_i^t\Phi_{i,y^2}\big)$, $i\!\!\in\!\!\{q,\varsigma\}$.
\end{theo}
\begin{IEEEproof}
	The proof is given in Appendix \ref{App4}.
\end{IEEEproof}
Eq.~(\ref{theo4}) specifies  the deterministic GML and includes a finite-range integral that can be evaluated numerically. The normal vector of the LS plane and the normal vector of the lens plane may lie in different  planes, which is referred to as ``out-of-plane reflection'' \cite{outofreflection}.  This is  in contrast to Snell's law which states that the reflected and the incident beam are in the same plane with $\theta_\ell=\theta_p$. However, depending on the chosen  phase shift profile across the IRS, the  direction of the reflected beam can be outside the incident beam plane.
Moreover, (\ref{theo4}) characterizes the dependence of the channel gain on the tile size, $L_x$ and $L_y$,  the phase shift profile across the tiles, $\Phi_q({\mathbf{r}},\mathbf{r}_q^t)$, the lens radius, $a$, and the lens and LS positions and orientations, $\mathbf{r}_\ell, \mathbf{\Psi}_\ell$ and $\mathbf{r}_p, \mathbf{\Psi}_p$, respectively.
Furthermore,  variables $A_0$ and $B_0$ defined in Theorem \ref{Theorem1} depend on the position of the center of the beam footprint on the IRS surface, $x_{\ell0}$ and $y_{\ell0}$, and the position where the  normal vector of the lens intersects the IRS, $x_{p0}$ and $y_{p0}$, which in turn  affects the GML and the end-to-end channel gain. In other words, to maximizes the received power due to the non-uniform power distribution across the IRS, the positions of the beam footprint and the lens center on the IRS should coincide. However, in case of misalignment, i.e.,  when the LS or the lens are not accurately tracked, only a fraction of the total power is received. We denote the corresponding misalignment error  by $\mathbf{r}_{e}=|\mathbf{r}_{\ell 0}-\mathbf{r}_{p0}|$ and investigate its impact on performance in Section \ref{Sec_Sim}.

In the following corollary, we simplify (\ref{theo4}), for the case where the normal vector of the LS plane and the normal vector of the lens plane lie in the same plane, e.g., the LS and the lens are at the same height, which is referred to as ``in-plane-reflection'' \cite{outofreflection}.
\begin{corol}[In-Plane Reflection]
	For in-plane reflection,  $\phi_p=\pi-\phi_\ell$ holds, and the deterministic GML  simplifies to
	\begin{IEEEeqnarray}{rll}
		h^\text{in}_\text{irs}&= \frac{C_h \pi}{4\sqrt{\rho_x\rho_y}} \sum\limits_{q=1}^Q\!\sum\limits_{\varsigma=1}^Q C_q C_\varsigma^* C_{2,q} C_{2,\varsigma}^*\exp\left(\frac{\varrho_x^2}{4\rho_x}+\frac{\varrho_y^2}{4\rho_y}\right)\nonumber\\ 
		&\times \left[\text{erf}\left(\sqrt{\rho_x}\tilde{a}+\frac{\varrho_{x}}{2\sqrt{\rho_x}}\right)-\text{erf}\left(-\sqrt{\rho_x}\tilde{a}+\frac{\varrho_{x}}{2\sqrt{\rho_x}}\right)\right]\nonumber\\
		&\times \left[\text{erf}\left(\sqrt{\rho_y}\tilde{a}+\frac{\varrho_{y}}{2\sqrt{\rho_y}}\right)-\text{erf}\left(-\sqrt{\rho_y}\tilde{a}+\frac{\varrho_{y}}{2\sqrt{\rho_y}}\right)\right]. \,\qquad
		\label{col}
	\end{IEEEeqnarray}
\end{corol}
\begin{IEEEproof}
	 Since we previously assumed $\phi_\ell=0$,  $\phi_p=\pi$ holds, and the parameters in Theorem \ref{Theorem4}  simplify to	 $\rho_x=\frac{k^2}{2d_p^2\tilde{b}_x}\sin^2(\theta_p)$, $\rho_y=\frac{k^2}{2d_p^2\tilde{b}_y}$,  $\rho_{xy}=0$, $\varrho_{x}=\frac{k^2\sin(\theta_p)}{2d_p}\left(\frac{{A}_{q}}{b_x}+\frac{{A}_{\varsigma}^*}{{b}_x^*}\right)$, and $\varrho_{y}=\frac{k^2}{2d_p}\left(\frac{{B}_{q}}{b_y}+\frac{{B}_{\varsigma}^*}{{b}_y^*}\right)$. Substituting these values, the integral in   (\ref{ptheo43})  simplifies  to	 two independent integrals, which can be solved by applying  \cite[Eq.~(2.33-1)]{integral}. Then, we obtain (\ref{col}) and this completes the proof.
\end{IEEEproof}
The results in Theorem \ref{Theorem1} and \ref{Theorem4} are valid for a given  IRS-based FSO link. To apply the above results to the IRS-based FSO link between the $m$-th LS and the $n$-th PD,  we  add the index of the LS $(m)$ and the index of the PD $(n)$ to all variables with index $(\ell)$ and $(p)$, respectively.
Thus, we obtain the deterministic GML, $h_\text{irs}^{m,n}$.  Based on our analysis in this section, we design optical  IRSs for  multi-link FSO systems in the following section. 

\section{IRS-Assisted Multi-Link FSO Systems}\label{Sec_Multilink}
\begin{figure}
	\centering
	\includegraphics[width=0.7\textwidth]{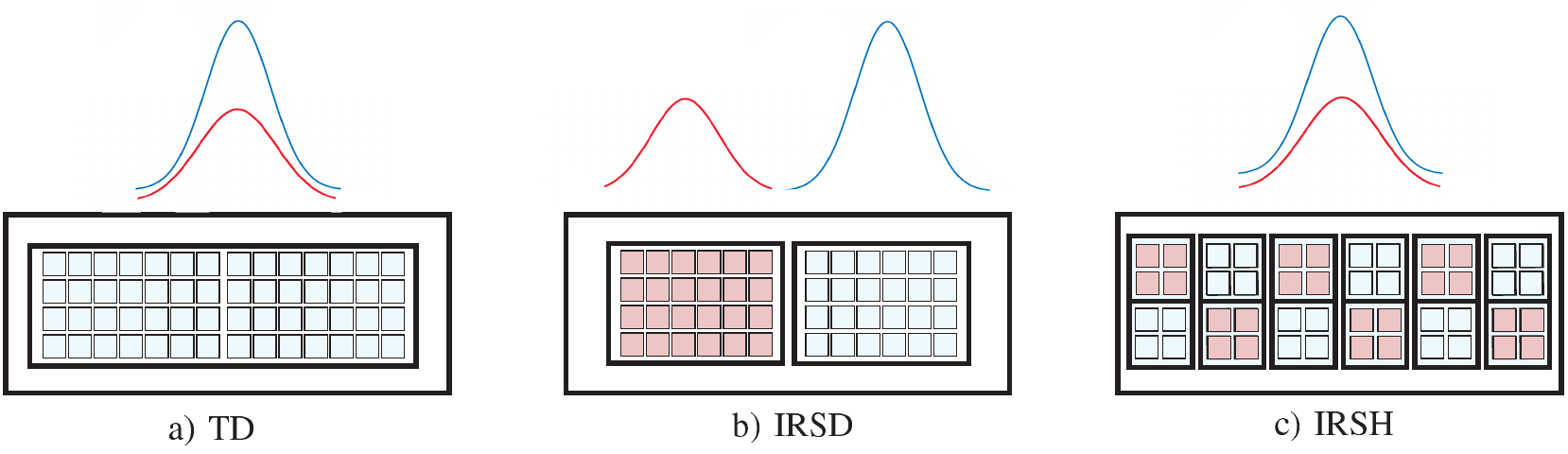}
	\caption{Illustration of three protocols for sharing the IRS between two LSs emitting Gaussian beams. The blue and red colors indicate the LS beams and the tiles allocated to the different LS-PD pairs.}
	\label{Fig:TileSelect}\vspace{-0.3cm}
\end{figure}

In the following,  we assume  multiple LSs  are connected via IRS-assisted FSO links to multiple PDs. First, we propose three  protocols  for  sharing the IRS surface  between multiple FSO links, namely the TD, IRSD, and IRSH protocols, see  Fig.~\ref{Fig:TileSelect}. These protocols  differ in their susceptibility to misalignment errors, the introduced delay, and the achieved signal-to-noise ratio (SNR), as will be shown in Section \ref{Sec_Sim}.  For each protocol, we first  specify  the required number of tiles, $Q$, the location of the center of the LS beam footprint  on the IRS, $\mathbf{r}_{\ell 0}$, and the location of the lens center on the IRS, $\mathbf{r}_{p0}$. Next, we assign the tiles to  LS-PD pairs and design  the tile parameters including  the phase shift profile, $\Phi_q(\mathbf{r},\mathbf{r}_q^t)$, and the efficiency factor, $\zeta_q$, as functions of the incident beam width, IRS size, and the IRS-lens  distance. 

\subsection{IRS Sharing Protocols} \label{Sec_IRSdesigns} 
\subsubsection{TD Protocol}
For the TD protocol, in each time slot, one LS  transmits, while the other LSs are inactive, thus, the  number of time slots, $T$, is identical to the number of FSO links, $N$, i.e.,  $T=N$. In each time slot, the entire IRS surface is  configured  for the active LS-PD pair, i.e., only one tile is needed and $Q=1$, see  Fig.~\ref{Fig:TileSelect}a).  Moreover, in order to maximize the received power, the phase shift profile center, the incident beam footprint center, and  the lens center  on the IRS  should  coincide with the origin, i.e., $\mathbf{r}_q^t=\mathbf{r}_{\ell 0,m}=\mathbf{r}_{p0,n}=(0,0,0),\, \forall m,n\in\mathcal{N}$. For this protocol, the active PD receives maximum power as the IRS serves only one LS-PD pair at a time.	However, the time sharing  among LSs degrades the achievable  data rate.
\subsubsection{IRSD Protocol} For this protocol, all LSs   simultaneously illuminate the IRS, which is divided into $Q=N$ tiles, see  Fig.~\ref{Fig:TileSelect}b). We assume  that each LS-PD pair is assigned to a different tile  such that the centers of the beam footprints, the lens centers, and the tile phase shift profile centers coincide with the centers of the respective tiles, i.e., $\mathbf{r}_{\ell 0,m}=\mathbf{r}_{p 0,n}=\mathbf{r}_q^t=\mathbf{r}_q, \forall m,n\in\mathcal{N}$. Since all LSs transmit simultaneously, the data rate may be increased compared to the TD protocol.  However,  due to the partitioning of the IRS among multiple FSO links, unless the IRS is sufficiently large, only part of the incident power is reflected by the respective tile towards the desired lens. Thereby,  the larger the incident beamwidth, the  less  power is received at the lens. Moreover, misalignment errors may shift the beam footprint center or the  lens center on the IRS  towards a tile reserved for a different LS-PD pair. Thus, given the Gaussian beam intensity distribution, a portion of the  power  may be  redirected  in an undesired direction which in turn degrades the GML. 
\subsubsection{IRSH Protocol} Since perfect tracking of the tile, beam footprint, and lens centers may not always be feasible, we propose the homogenization of the IRS surface by dividing it  into  $Q$ tiles such that the number of tiles is much larger than the number of LS-PD pairs, i.e., $Q\gg N$, see  Fig.~\ref{Fig:TileSelect}c).  Unlike for the IRSD protocol, where one tile is assigned to one LS-PD pair, for this protocol, multiple smaller tiles are allocated to each LS-PD pair, and thus, the LS beam power is distributed across multiple tiles. Thus, if  the beam footprint center or/and the lens center on the IRS are shifted due to misalignment errors, the impact on the receive power will be mitigated. This benefit comes at the cost of  a power loss since only  some of  the tiles around beam center are designed to redirect the beam towards the desired lens. Thereby, the IRSH protocol  is more resilient  to    misalignment errors, but without misalignment,   the IRSD and TD protocols may achieve a higher performance as they redirect more power to the desired lenses.
\subsection{Tile Configuration}\label{Sec_Tile_Config}
For all proposed IRS sharing protocols, at a given time, each  tile is assigned to only one LS-PD pair.  In the following, we determine the phase shift profile of the $q$-th tile, $\Phi_q(\mathbf{r},\mathbf{r}_q^t)$,  needed to redirect  the LS beam towards a desired PD and the corresponding efficiency factor, $\zeta_q$, required for a passive-lossless IRS.
\subsubsection{Tile Phase Shift Profile $\Phi_q(\mathbf{r},\mathbf{r}_q^t)$} The  phase shift introduced by the tile is exploited such that it compensates the phase difference between the incident and the desired beam.  Moreover,  the  incident beam on the IRS has a nonlinear phase profile (\ref{lem1}) and the phase profile of the secondary waves (\ref{Huygens-Fresnel}) include linear or quadratic terms depending  on the operating regime and the approximation used, cf. Section \ref{Sec_IntvsFar}. Thus, the desired phase shift across the tile is in general a nonlinear function. Assuming that the positions of the $m$-th LS, $\left(d_{\ell m}, \mathbf{\Psi}_{\ell m}\right)$, and the $n$-th PD, $\left(d_{pn},\mathbf{\Psi}_{pn}\right)$, are known, we adopt LP and QP profiles as  approximations of  general nonlinear phase shift profiles as follows
\begin{IEEEeqnarray}{rll}\label{Phase-shift}
		\Phi_{q}^\text{quad}(\mathbf{r},\mathbf{r}_q^t)&=k\left(\Phi_{q,0}+\Phi_{q,x} (x-x_q^t)+\Phi_{q,y} (y-y_q^t)+\Phi_{q,x^2} (x-x_q^t)^2+\Phi_{q,y^2} (y-y_q^t)^2\right),	\qquad\IEEEyesnumber\IEEEyessubnumber \label{QP}\\
		\Phi_{q}^\text{lin}(\mathbf{r},\mathbf{r}_q^t)&=k\left(\Phi_{q,0}+\Phi_{q,x} (x-x_q^t)+\Phi_{q,y} (y-y_q^t)\right),\qquad\IEEEyessubnumber\label{LP}
\end{IEEEeqnarray}
where  $\Phi_{q,x}=\cos(\theta_{\ell m})\cos(\phi_{\ell m})+\cos(\theta_{pn})\cos(\phi_{pn})$,  $\Phi_{q,y}=\cos(\theta_{\ell m})\sin(\phi_{\ell m})+\cos(\theta_{pn})\sin(\phi_{pn})$,  $\Phi_{{q,x^2}}=\frac{1}{2d_{pn}}\left(1+\cos^2(\theta_{pn})\cos^2(\phi_{pn})\right)-\frac{\sin^2(\theta_{\ell m})}{2R(\hat{d}_{\ell m})}-\frac{1}{4d_{pn}}$,  $\Phi_{q,y^2}\!=\frac{1}{2d_{pn}}\big(1\!+\cos^2(\theta_{pn})\sin^2(\phi_{pn})\big)-\frac{1}{2R(\hat{d}_{\ell m})}-\frac{1}{4d_{pn}}$, and $\Phi_{q,0}=d_{pn}-\hat{d}_{\ell m}$.  The LP profile is chosen such that the total accumulated phase, see (\ref{phas-tot}), is zero when the beam arrives at the lens center. Then,  for the QP profile, the quadratic terms cancel the total accumulated phase due to the LS beam curvature and  the IRS-to-lens distance. Furthermore, adding  the term $-\frac{1}{4d_{pn}}$  causes  a parabolic phase profile that focuses the beam at the lens center. These phase profiles  allow the application of Theorems \ref{Theorem1} and \ref{Theorem4}. Moreover, the  LP profile in (\ref{LP})  corresponds to the generalized Snell's law given in \cite{genralized-snell, Marzieh_IRS_jou}. 
\subsubsection{Tile Efficiency $\zeta_q$} The efficiency factor $\zeta_q$  ensures that the power reflected from the tile is smaller  or equal to the  power incident on the tile. We assume $\zeta_q=\zeta_0\bar{\zeta}_q$, where $\zeta_0$ denotes the resistive loss of the IRS and $\bar{\zeta}_q$ is the passivity constant which ensures the incident power is  completely reflected. 
\begin{prop}\label{Passivity}
	The passivity factor $\bar{\zeta}_q$ of tile $q$  which is assigned to the $m$-th LS and the $n$-th lens depends on $\mathbf{\Psi}_{pn}$ and is given by
	\begin{IEEEeqnarray}{rll}
		\bar{\zeta}_q=\sqrt{\lvert\sin(\theta_{p n})\rvert}.
		\label{zeta}
	\end{IEEEeqnarray}	
\end{prop}  
\begin{IEEEproof}
	The proof is given in Appendix \ref{App5}.
\end{IEEEproof}
In particular,  using the above result for anomalous reflection in the far-field scenario in (\ref{ellips_Gaus}), we obtain  ${\zeta}_t^2=\frac{\lvert\sin(\theta_{p n})\rvert}{\lvert\sin(\theta_{\ell m})\rvert}$, which is similar to the result for  plane waves given in \cite{Estakhri,Vahid_IRS}.
\section{Performance Analysis}\label{Sec_Perf}
 For the IRSD and IRSH protocols, the IRS surface is shared by multiple links. Hence, in principle,  inter-link interference can affect the end-to-end performance. However,  by careful positioning of the LSs and PDs, inter-link interference can be considerably  mitigated. Nevertheless, in the following, we analyze the end-to-end performance of the considered IRS-assisted multi-link FSO transmission systems  taking into account the possibility of inter-link interference. 
\subsection{Bit Error Rate}
To analyze the average BER for the $n$-th LS-PD pair,   we first derive the instantaneous BER for a given realization of the atmospheric turbulence induced fading in the presence of interference in the following Lemma.
\begin{lem}\label{remk_BER}
Assuming OOK modulated symbols $s_m$ with signal power $P_{\ell m}, m\in\mathcal{N}$, and AWGN noise $w_n\sim\mathcal{N}(0,\sigma_w^2)$,  an error occurs if   symbol $\hat{s}_n\neq s_n$ is detected by the $n$-th PD. The corresponding BER is given by
\begin{IEEEeqnarray}{rll}
	P_{e,n}(\mathbf{h}_a^{n})&=\frac{1}{2^{N}}\sum_{\mathbf{s}_m\in\mathcal{S}_m}Q\left(\frac{1}{2}h_a^{n,n}\sqrt{\gamma_n}-\!\sum_{\substack{m\in\mathcal{N}\\ m\neq n}}h_a^{m,n}s_m\sqrt{\gamma_{m}}\right)\nonumber\\
	&+\frac{1}{2^{N}}\sum_{\mathbf{s}_m\in\mathcal{S}_m}Q\left(\frac{1}{2}h_a^{n,n}\sqrt{\gamma_n}+\!\sum_{\substack{m\in\mathcal{N}\\ m\neq n}}h_a^{m,n}s_m\sqrt{\gamma_{m}}\right),
	\label{PEP}
\end{IEEEeqnarray}
where $\mathbf{h}_a^{n}=[h_a^{1,n}, \dots, h_a^{N,n}]$, $\mathbf{s}_m=[s_1,\dots, s_{n-1},s_{n+1},\dots, s_N]$ denotes the vector of interfering symbols $s_m$,   $\mathcal{S}_m=\{0,1\}^{N-1}$, $\gamma_\iota=\frac{P_{\ell \iota}}{\sigma_w^2}|h_{\text{irs}}^{\iota,n}h_p^{\iota,n}|^2, \iota\in\{n,m\}$, and $Q(\cdot)$ is the Gaussian Q-function.
\end{lem}
\begin{IEEEproof}
	The proof is given in Appendix \ref{App6}.
\end{IEEEproof}
 Then, given that $h_a^{m,n}, \forall m\in\mathcal{N},$  are independent random variables,  the average  BER for the $n$-th LS-PD pair, denoted by $P_{e,n}$, is given by 
 \begin{IEEEeqnarray}{rll}
 	P_{e,n} =\int_{0}^{\infty} \dots\int_{0}^{\infty} P_{e,n}(\mathbf{h}_a^{n}) f_{h_a^{1,n}}(h_a^{1,n})\dots f_{h_a^{N,n}}(h_a^{N,n}) \mathrm{d}\mathbf{h}_a^{n},
 	\label{BER}
 \end{IEEEeqnarray}
where    $f_{h_a^{m,n}}(\cdot)$ is the probability density function (PDF) of a   Gamma-Gamma distributed random variable which is given by \cite{Krog_sum_GG}
\begin{IEEEeqnarray}{rll}
	f_{h_a^{m,n}}(h_a^{m,n})=\frac{2(\alpha_{m,n}\beta_{m,n} h_a^{m,n})^{\frac{\alpha_{m,n}+ \beta_{m,n}}{2}}}{\Gamma(\alpha_{m,n})\Gamma(\beta_{m,n})h_a^{m,n}} K_{\alpha_{m,n}-\beta_{m,n}}\left(2\sqrt{{\alpha_{m,n}\beta_{m,n}} h_a^{m,n}}\right),
	\label{GG}
\end{IEEEeqnarray}
where  $\Gamma(\cdot)$ is the Gamma function, and $K_v(\cdot)$ is the modified Bessel function of the second kind with order $v$.  Eq.~(\ref{BER}) involves an $N$ dimensional integral which can be  evaluated numerically, especially if $N$ is small, e.g.,   for $N=2$ LS-PD pairs, as considered  in Section \ref{Sec_Sim} for our numerical results. For large $N$, (\ref{BER}) can be  approximated using the PDF of the sum of Gamma-Gamma distributed variables in \cite[Eqs.~(24) and (25)]{Krog_sum_GG}.

For the  special case of noise limited systems, where the inter-link interference in (\ref{PEP}) can be ignored, we obtain $P_{e,n} = P_e\left(\sqrt{\frac{\gamma_n}{4}}\right)$ \cite{Error-Robert}, where $P_e(x)$ is given by
\begin{IEEEeqnarray}{rll}
	P_{e}(x) =& \sum_{\iota=0}^{\infty}\left(\xi_{\iota}(\alpha_{n,n}, \beta_{n,n})(4x)^{-{\iota+\beta_{n,n}\over 2}}+\xi_{\iota}(\beta_{n,n}, \alpha_{n,n})(4x)^{-{\iota+\alpha_{n,n}\over 2}}\right),
\end{IEEEeqnarray}
and 
\begin{IEEEeqnarray}{rll}
	&\xi_{\iota}(\alpha_{n,n}, \beta_{n,n}) ={\sqrt{\pi}(2\sqrt{2}\alpha_{n,n}\beta_{n,n})^{\iota+\beta_{n,n}}{\Gamma}\left({\iota+\beta_{n,n}+1\over 2}\right)\over 2\sin[\pi(\alpha_{n,n}-\beta_{n,n})]\Gamma(\alpha_{n,n})\Gamma(\beta_{n,n}) \Gamma(\iota-\alpha_{n,n}+\beta_{n,n}+1)(\iota+\beta_{n,n})\iota!}. \qquad
\end{IEEEeqnarray}

\subsection{Outage Probability}
Given the high data rates  of  FSO systems (10 $\text{Gbit}/\text{s}$) compared to the coherence time  of the channel (1-10 ms) and in order to investigate the trade-off between the transmission rate and the received  power of the proposed IRS protocols,   we study the outage probability.  The  outage probability of the $n$-th LS-PD pair  is defined as the probability that the fixed data rate of the LS, $R_n$, exceeds the capacity of the end-to-end channel, $\mathcal{C}_n$, and is given by
\begin{IEEEeqnarray}{rll}
	P_{\text{out},n}=  \text{Pr}\{\mathcal{C}_n<R_n\}.
	\label{Pout}
\end{IEEEeqnarray}
To evaluate (\ref{Pout}), first we provide  a lower bound for the capacity of the FSO interference channel in the following lemma.  
\begin{lem}\label{remk_CAP}
Assuming $s_n[t]\in\mathbb{R}^+$ with an average  power constraint, $\mathbb{E}\{|s_n[t]|^2\}\leq P_{\ell n}$, and Gaussian noise $w_n[t]$ with variance $\sigma_{w}^2$, the capacity of the interference channel in (\ref{system_eq}) is lower bounded by
\begin{IEEEeqnarray}{rll}
		\mathcal{C}_n\geq  \mathcal{C}_{\text{low}}(\Upsilon_n),
		\label{Interf_Cap}
\end{IEEEeqnarray}
where $\mathcal{C}_{\text{low}}(\Upsilon_n)=\frac{W_\text{FSO}}{2} \ln\left(1+\frac{\Upsilon_n e}{2\pi}\right)$, $\Upsilon_n=\frac{|h_{n,n}|^2P_{\ell n}}{\tilde{\sigma}_s^2+\sigma_{{w}}^2}$, and $\tilde{\sigma}_{s}^2=\sum\limits_{\substack {m=1\\ m\neq n}}^N P_{\ell m} |h_{m,n}|^2$.
\end{lem}	
\begin{IEEEproof}
		The proof is given in Appendix \ref{App7}.
\end{IEEEproof} 
Using the above lemma and (\ref{Pout}), we obtain an upper bound on the  outage probability in (\ref{Pout}) as follows	
 \begin{IEEEeqnarray}{rll}
	P_{\text{out},n}\leq P_{\text{out},n}^{\text{up}}=  \text{Pr}\{\mathcal{C}_{\text{low}}(\Upsilon_n)< R_n\}\overset{(a)}{=}	\text{Pr}\{\Upsilon_n< \gamma_\text{thr}\},
	\label{Pout_lower}
\end{IEEEeqnarray}
where for  $(a)$, we exploit that $\mathcal{C}_{\text{low}}(\cdot)$ is a  monotonically increasing function in $\Upsilon_n$, and thus, an outage  occurs when the received SINR falls below threshold $\gamma_\text{thr}=\mathcal{C}_{\text{low}}^{-1}\left(R_n\right)=\frac{2\pi}{e}\left(e^{2R_n/ W_\text{FSO}}-1\right)$. Next, given the randomness of the turbulence induced fading, $h_a^{m,n}$,  $P_{\text{out},n}^{\text{up}}$ is given by
\begin{IEEEeqnarray}{rll}
&P_{\text{out},n}^{\text{up}}=  \text{Pr}\{\Upsilon_n< \gamma_\text{thr}\}=\text{Pr}\left\{\left(h_a^{n,n}\right)^2< \bar{\chi}\right\}=\nonumber\\
&\int_{0}^{\infty} \dots\int_{0}^{\infty} F_{h_a^{n,n}}\left(\sqrt{\bar{\chi}}\right) f_{h_a^{1,n}}(h_a^{1,n})\dots f_{h_a^{n-1,n}}(h_a^{n-1,n})f_{h_a^{n+1,n}}(h_a^{n+1,n})\dots f_{h_a^{N,n}}(h_a^{N,n}) \mathrm{d}\bar{\mathbf{h}}_a^{n},\qquad
\label{Pout_last_general}
\end{IEEEeqnarray}
where  $\bar{\mathbf{h}}_a^{n}=[h_a^{1,n}, \dots, h_a^{n-1,n}, h_a^{n+1,n}, \dots h_a^{N,n}]$, $\bar{\chi}={\gamma_\text{thr}\over\gamma_{n}}\sum\limits_{\substack{m=1\\ m\neq n}}^N  \gamma_m |h_a^{m,n}|^2+{\gamma_\text{thr}\over\gamma_{n}}$, and $F_{h_a^{n,n}}(\cdot)$ denotes the cumulative distribution function (CDF) of the Gamma-Gamma distribution given in \cite[Eq.~(7)]{cdf_GG}. For small $N$, the above equation can be numerically integrated. For large $N$, (\ref{Pout_last_general}) can be  approximated using the PDF of the sum of squared Gamma-Gamma distributed variables given in \cite[Eq.~(9)]{cdf_Inft}. 

As a special case, for noise limited systems, where the impact of inter-link interference can be ignored, the upper bound on the outage probability can be simplified as follows
\begin{IEEEeqnarray}{rll}
	P_{\text{out},n}^{\text{up}}= \text{Pr}\{\left(h_a^{n,n}\right)^2\leq \Upsilon_\text{thr}\}=F_{h_a^{n,n}}\left(\sqrt{\Upsilon_\text{thr}}\right),
	\label{Pout_last}
\end{IEEEeqnarray}
where $\Upsilon_\text{thr}=\frac{\gamma_\text{thr}}{\gamma_{n}}$. Substituting $F_{h_a^{n,n}}(\cdot)$ given in \cite[Eq.~(7)]{cdf_GG} in (\ref{Pout_last}), the upper bound on the outage probability is given by \cite{Alouini-GG}
\begin{IEEEeqnarray}{rll}
P_{\text{out},n}^{\text{up}} = {1 \over \Gamma(\alpha_{n,n}) \Gamma(\beta_{n,n})} \, G_{1, 3}^{2, 1}\left[\alpha_{n,n}\beta_{n,n}\sqrt{\Upsilon_\text{thr}}\bigg\vert_{\alpha_{n,n}, \beta_{n,n}, 0}^{1}\right],
\label{Outage_woInf}
\end{IEEEeqnarray}
where $G_{m, n}^{p, q}\left[x\bigg\vert_{b_1, \cdots, b_q}^{a_1, \cdots, a_p}\right]$ is the Meijer's G-function.
\section{Simulation Results}\label{Sec_Sim}
\begin{table}[t]
	\centering
	\caption{System and channel parameters \cite{WCNC},  \cite{Marzieh_IRS_jou}.}
	\scalebox{0.55}{%
		\begin{tabular}{|| l | c | c || }
			\hline  
			{LS Parameters} &{Symbol}& {Value}  \\ 
			\hline
			\hline
			FSO bandwidth &$W_{\text{FSO}}$ &$1\ \mathrm{GHz}$\\
			FSO wavelength &$\lambda$ &$1550\, \mathrm{nm}$\\
			Beam waist radius &$w_{01}, w_{02}$& $0.25\ \mathrm{mm}$\\
			Electric fields at origin &$E_{01},E_{02}$ &$60\, \frac{\mathrm{kV}}{\mathrm{m}}$\\
			Noise spectral density &$N_0$&$-114\ \mathrm{dBm}/\mathrm{MHz}$\\
			Attenuation coefficient &$\kappa$&$0.43\times 10^{-3} \frac{\mathrm{dB}}{\mathrm{m}}$\\
			Distance and orientation of LS 1  &$d_{\ell 1},  \mathbf{\Psi}_{\ell 1}$ &$1\, \text{km},\left({\pi\over 3}, 0\right)$\\
			Distance and orientation of LS 2  &$d_{\ell 2},  \mathbf{\Psi}_{\ell 2}$ &$1\, \text{km},\left({\pi\over 4}, 0\right)$\\ 
			Gamma-Gamma parameters &$(\alpha, \beta)$ &$(2, 2)$\\
			Impedance of the propagation medium &$\eta$ &$377 \, \Omega$\\
			\hline
			\hline
			{IRS Parameters}&& \\
			\hline 
			\hline
			IRS Size  &$L_{x,\text{tot}}\times L_{y,\text{tot}}$ & $1$ m $\times 0.5$ m\\
			Separation distance between tiles &$l_x, l_y$ &$0,0$\\
			Number of tiles for TD protocol &$Q_x\times Q_y$ &$1\times1$\\
			Number of tiles for  IRSD protocol &$Q_x\times Q_y$ &$2\times1$\\	
			Number of tiles for  IRSH protocol &$Q_x\times Q_y$ &$8\times2$\\			
			\hline
			\hline
			{PD and Lens Parameters}&&  \\
			\hline
			\hline
			Lenses radius &$a$&$15\ \mathrm{cm}$\\
			Distance and orientation of lens of PD 1 &$d_{p 1},  \mathbf{\Psi}_{p 1}$ &$3\, \text{km},\left({\pi\over 3}, \pi\right)$\\
			Distance and orientation of lens of PD 2 &$d_{p 2},  \mathbf{\Psi}_{p 2}$ &$3\, \text{km},\left({\pi\over 6}, \pi\right)$\\ 
			\hline
		\end{tabular}
	}
	\label{Table:Sys_Param}\vspace{-6mm}
\end{table}
In this section, first we validate the analytical channel gain in (\ref{theo4}) for a point-to-point IRS-assisted FSO link. Then, we consider the case where two LS-PD pairs share one IRS, see Fig.~\ref{Fig:Setup}, and determine the impact of inter-link interference.  Finally, we  investigate the system performance in terms of BER and outage probability.  The  LP  profile across the IRS in (\ref{LP}) is applied except  for Fig.~\ref{Fig:BER}, where also the QP profile in (\ref{QP}) is considered. For all figures, the parameter values provided in Table \ref{Table:Sys_Param} are adopted, unless specified otherwise. 

\subsection{Validation of the Channel Model}
\begin{figure}[t]
	\centering
	\begin{minipage}[h]{0.4\textwidth}
		\centering
		\includegraphics[width=0.9\textwidth]{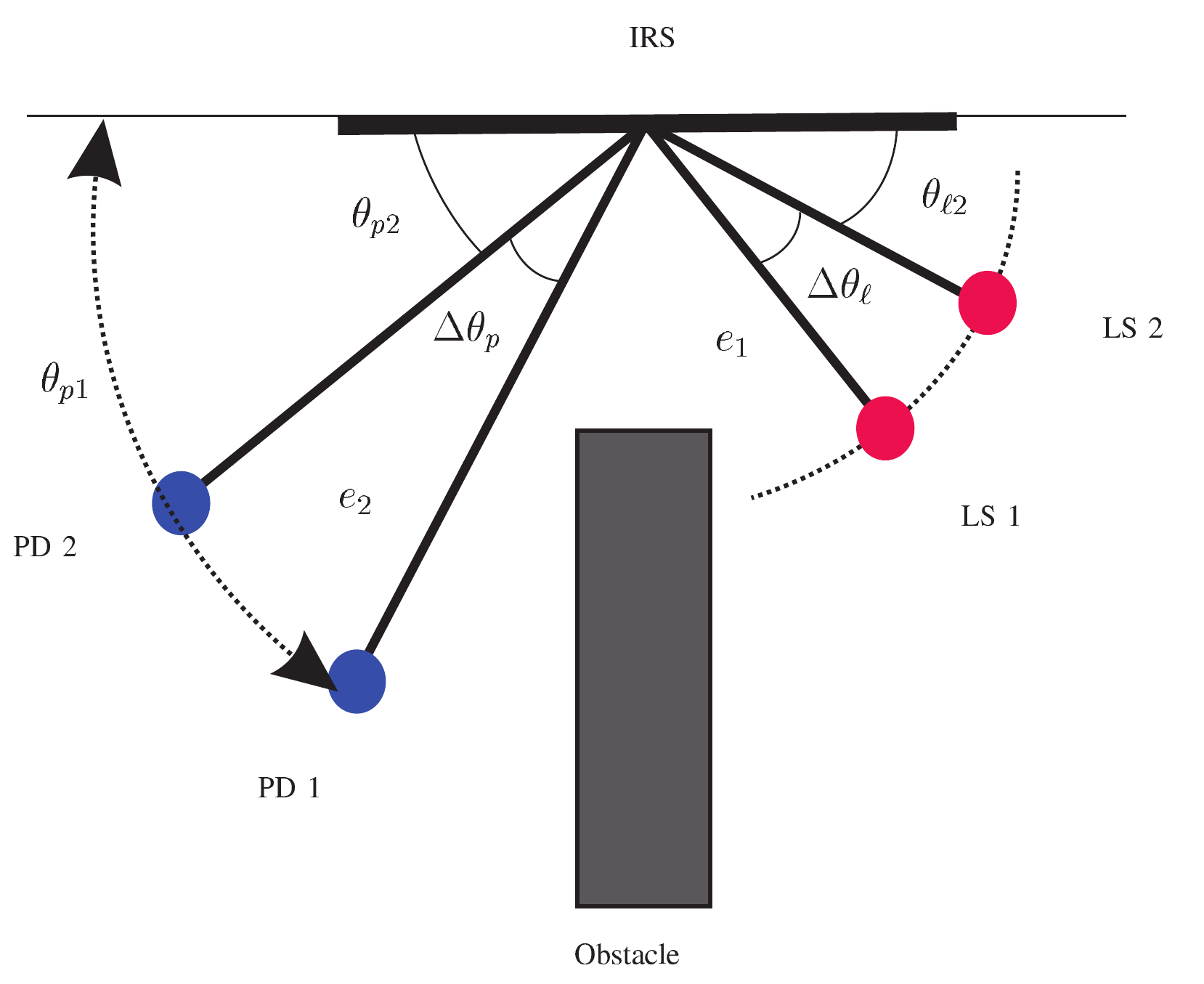}
		\vspace*{18mm}
		\caption{ Simulation setup.\label{Fig:Setup}}
		
	\end{minipage}
	\begin{minipage}[h]{0.55\textwidth}
		\centering
		\includegraphics[width=1\textwidth]{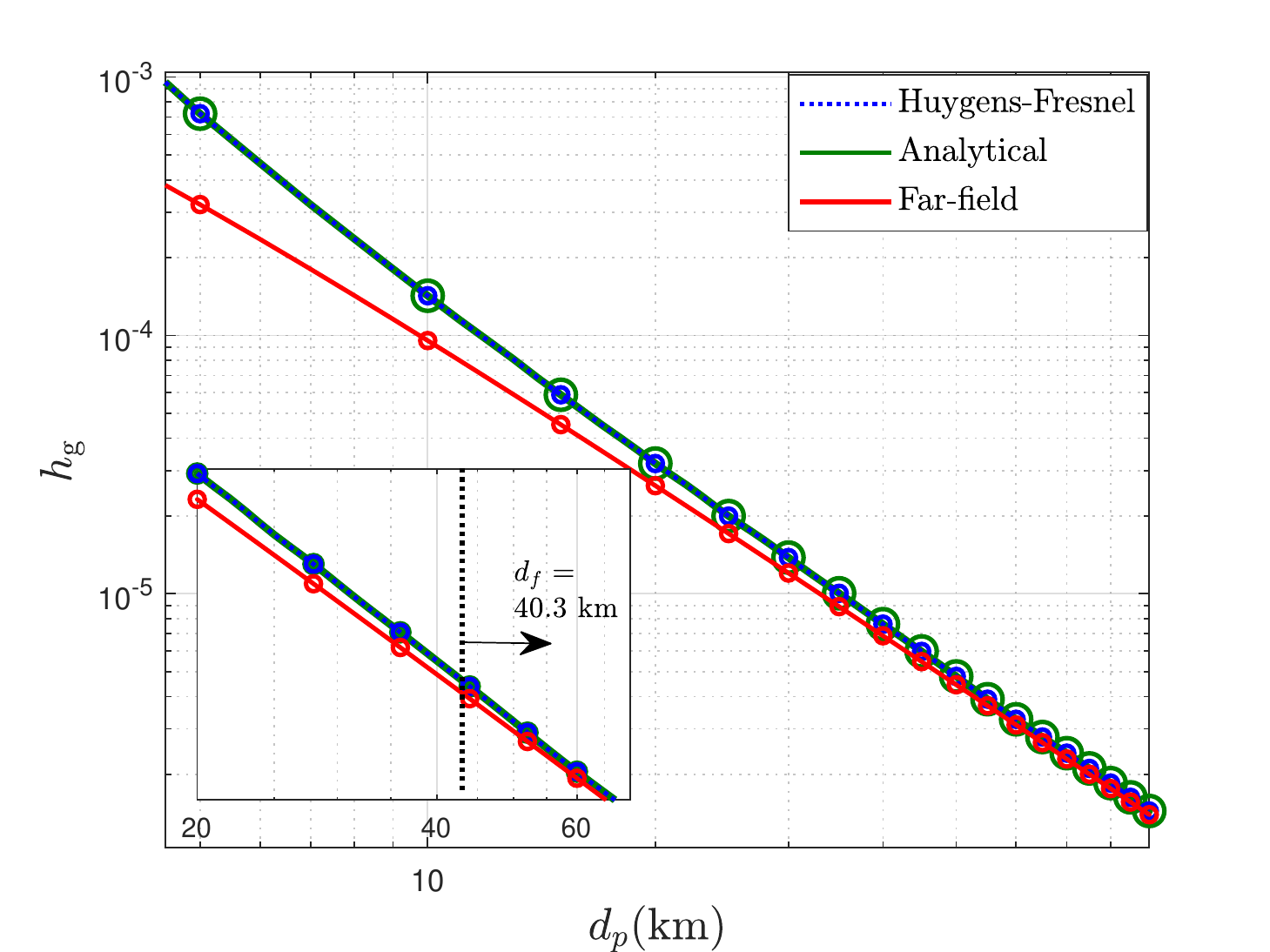}
		\caption{GML for IRS-based FSO channel between LS 1 and PD 1 versus $d_{p1}$.\label{Fig:hirs}}
	\end{minipage}
	\vspace*{-6mm}
\end{figure}

First, we  validate  our analytical results  for the point-to-point GML  for the IRS-assisted FSO system in (\ref{theo4}). For Fig.~\ref{Fig:hirs}, we assume that LS 1 is connected via a single-tile IRS to PD 1. Fig.~\ref{Fig:hirs} shows  the analytical GML, $h_\text{irs}^{1,1}$, in (\ref{theo4}),  the numerical integration of the Huygens-Fresnel integral  using (\ref{Huygens-Fresnel}), (\ref{irradiance}), and (\ref{gain}), and the  GML obtained  with the  far-field approximation in (\ref{ellips_Gaus}),  where an  IRS   size of $L_{x,\text{tot}}=L_{y,\text{tot}}=0.5$ m is assumed. As can be observed, for larger  IRS-lens distances, the GML decreases which is due to the divergence of the laser beam along the propagation path. Thus, for large distances,  the lens receives a smaller portion of the LS beam power which leads to a smaller  GML. Fig.~\ref{Fig:hirs} also shows that the proposed analytical GML in (\ref{theo4})  matches  the Huygens-Fresnel results for both IRS sizes, whereas  the  far-field approximation in  (\ref{ellips_Gaus}) is only valid in the far-field regime.  The subfigure in Fig.~\ref{Fig:hirs}  shows the far-field regime. The mismatch between the GML for the far-field approximation and  the Huygens-Fresnel results can be explained based on the definition of the far-field distance in (\ref{d_f}). Here, the LS beam width  incident on the IRS is $w_x(\hat{d}_{\ell 1})=2.28$ m and $w_y(\hat{d}_{\ell 1})=1.97$ m.  From  (\ref{d_f}), we obtain the minimum far-field distance  as $d_f=40.3$ km. At distances $d_{p1}>d_f$, the analytical GML and the Huygens-Fresnel results approach the far-field approximation.  Fig.~\ref{Fig:hirs}  confirms that for the typical range  of FSO systems  of a few kilometers, the proposed analytical channel model is valid for practical IRS sizes and IRS-lens distances, whereas  the far-field approximation does not always yield accurate results.

\subsection{Interference Channel Power}
\begin{figure}[t]
	\centering
	\begin{subfigure}[h]{0.495\textwidth}
		\centering
		\includegraphics[width=1\textwidth]{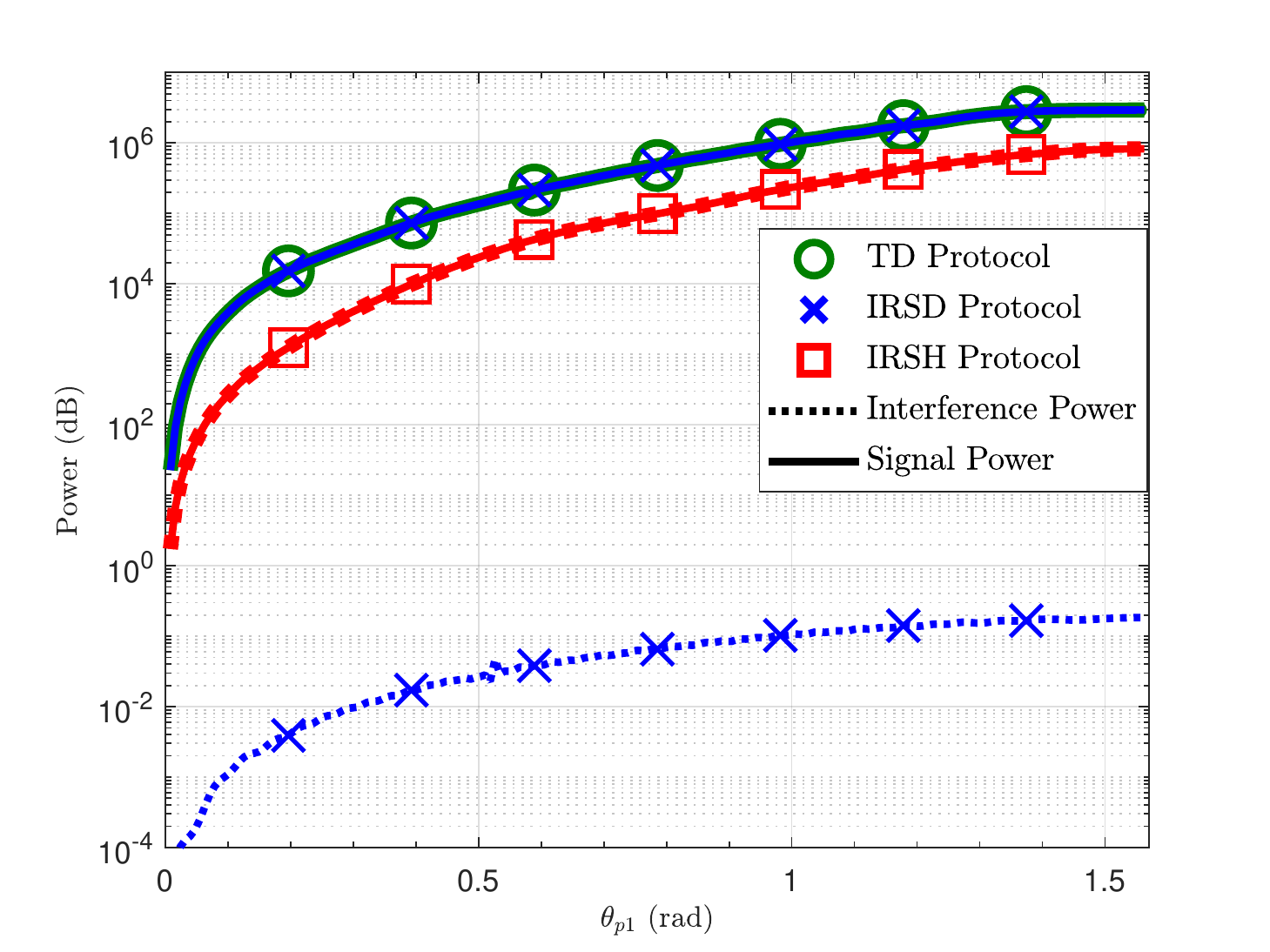}
		\vspace{-5mm}
		\caption{\label{Fig:IntfD0}} 
	\end{subfigure}
	\begin{subfigure}[h]{0.49\textwidth}
		\centering
		\includegraphics[width=1\textwidth]{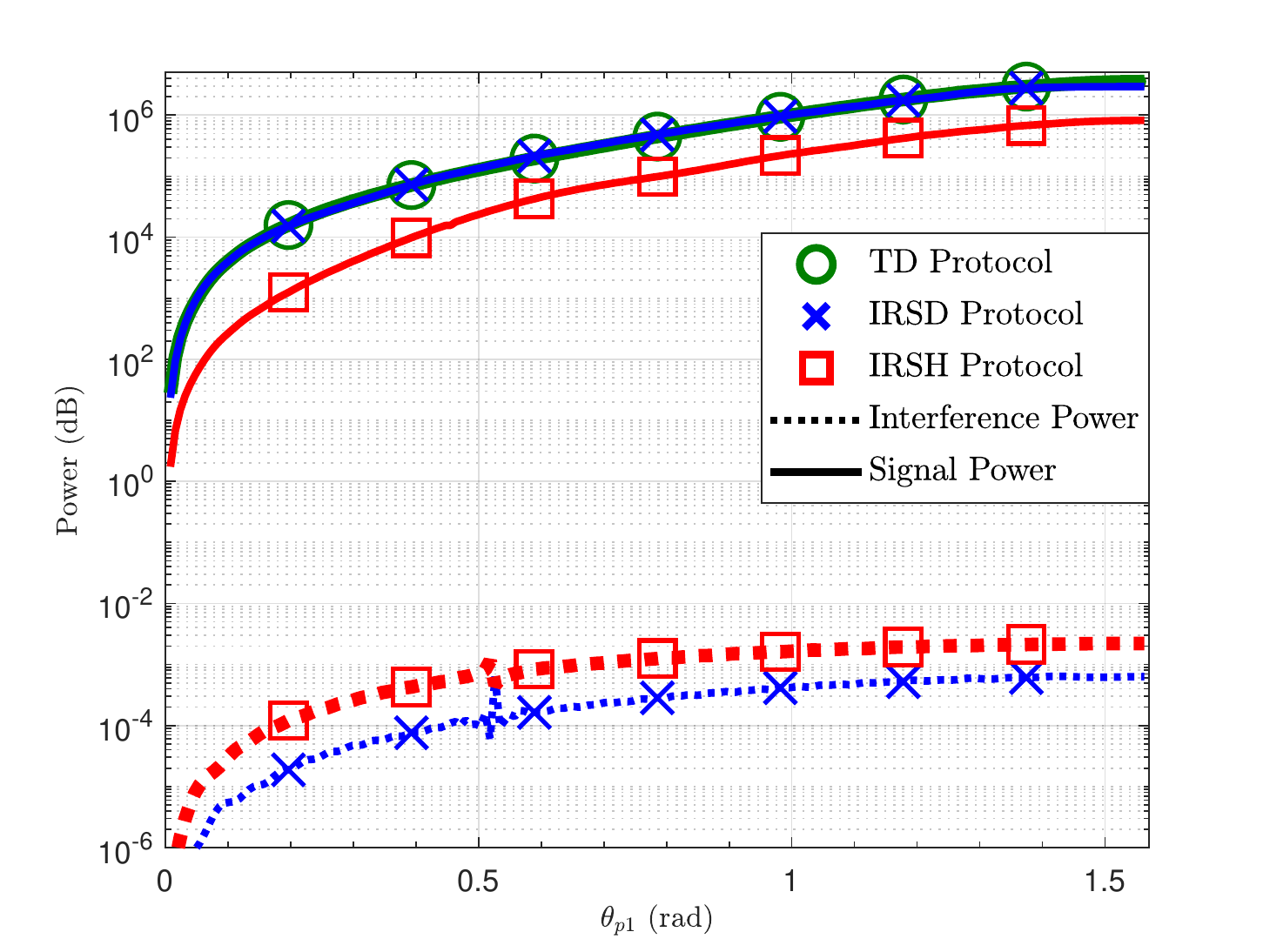} 
		\vspace{-5mm}
		\caption{\label{Fig:IntfD1}} 
	\end{subfigure}
	\vspace{-2mm}
	\caption {The power of the desired signal from LS 1 at PD 1 and the interference signal from LS 2 at PD 1. The figure shows the power versus $\theta_{p1}$  for two setups (a) $\Delta\theta_\ell=0$ rad and  (b) $\Delta\theta_\ell=1$ mrad.\label{Fig:Intf}}
	\vspace{-5mm}
\end{figure}

In the following, we consider a rectangular-shaped IRS of size  $L_{x,\text{tot}}=1$ m and $L_{y,\text{tot}}=0.5$ m. Two  LS-PD pairs are connected via the IRS, see Fig.~\ref{Fig:Setup} .  The LSs and lenses  are located on circles centered at the origin and with radii $e_1=1$ km and $e_2=3$ km, respectively. LS 2 and PD 2 are fixed at angles $\mathbf{\Psi}_{\ell 2}=({\pi \over 4}, 0)$ and  $\mathbf{\Psi}_{p 2}=({\pi \over 6}, \pi)$, respectively. LS 1 and PD 1 are located at angles $\mathbf{\Psi}_{\ell 1}=({\pi \over 4}+\Delta\theta_\ell, 0)$ and  $\mathbf{\Psi}_{p 1}=(\theta_{p1}={\pi \over 6}+\Delta\theta_p, \pi)$, respectively, i.e., the two LSs and the two PDs are separated by angles   $\Delta\theta_\ell$ and $\Delta\theta_p$, respectively. The IRS employs the TD, IRSD, and IRSH  protocols to allow both LS-PD pairs to share its surface.  PD 1  receives the reflected signal of the desired transmitter  LS 1 and for the IRSD and IRSH protocols, that of the interfering transmitter LS 2. Fig.~\ref{Fig:Intf} shows the signal  power, $\gamma_1$,  and the interference power, $\gamma_2$, received  at PD 1 as a function of angle $\theta_{p1}$.  Results for LS  separation angles  $\Delta\theta_\ell=0$ mrad and 1 mrad  are shown in Figs. \ref{Fig:IntfD0} and \ref{Fig:IntfD1}, respectively. As can be observed,  by increasing $\theta_{p1}$, both the received signal power and the interference power  increase since the LS power captured by the lens plane is maximized when it is parallel to the IRS plane. 
Moreover, Figs.~\ref{Fig:IntfD0} and \ref{Fig:IntfD1} reveal  that the TD  and  IRSD protocols capture the same amount of signal power since for both protocols the respective IRS tile  is smaller than the  beam footprint in the IRS plane.  Furthermore, for the  TD and IRSD protocols,  the beam axis of LS 1  and the lens center on the IRS  coincide with  the   center of the tile configured for LS 1 and PD 1,  whereas   the IRSH protocol disregards the  positioning of the beam axis and lens centers on the tile. Thus, PD 1 can collect more power  for the TD and IRSD  protocols than for the IRSH  protocol.  Moreover, for the IRSD protocol, the interference power for both values of $\Delta\theta_\ell$ is considerably smaller than the signal power since   the two LS beams  and the two lens centers are assigned to the centers of different IRS tiles. For the IRSH  protocol, the interference power and  the signal power  are similar when the LSs are co-located ($\Delta\theta_\ell=0$ mrad), see Fig.~\ref{Fig:IntfD0}, whereas after slightly increasing the separation angle between the LSs to $\Delta\theta_\ell=1$ mrad in Fig.~\ref{Fig:IntfD1}, the IRSH protocol yields a similar interference power as  the IRSD protocol. In summary, for the IRSD and IRSH protocols, the interference can be limited by suitable positioning of the beam footprints   and the lens centers on the IRS and by careful positioning of the LSs and PDs, respectively. 

\subsection{Performance Analysis}

In the following, we investigate the performance of the setup specified  in Fig.~\ref{Fig:Setup} and Table \ref{Table:Sys_Param} in terms of  BER and outage probability. The simulation results are averaged over $10^6$ channel realizations.

\begin{figure}[t!]
	\centering
	\begin{subfigure}[h]{0.48\textwidth}
		\centering
		\includegraphics[width=1\textwidth]{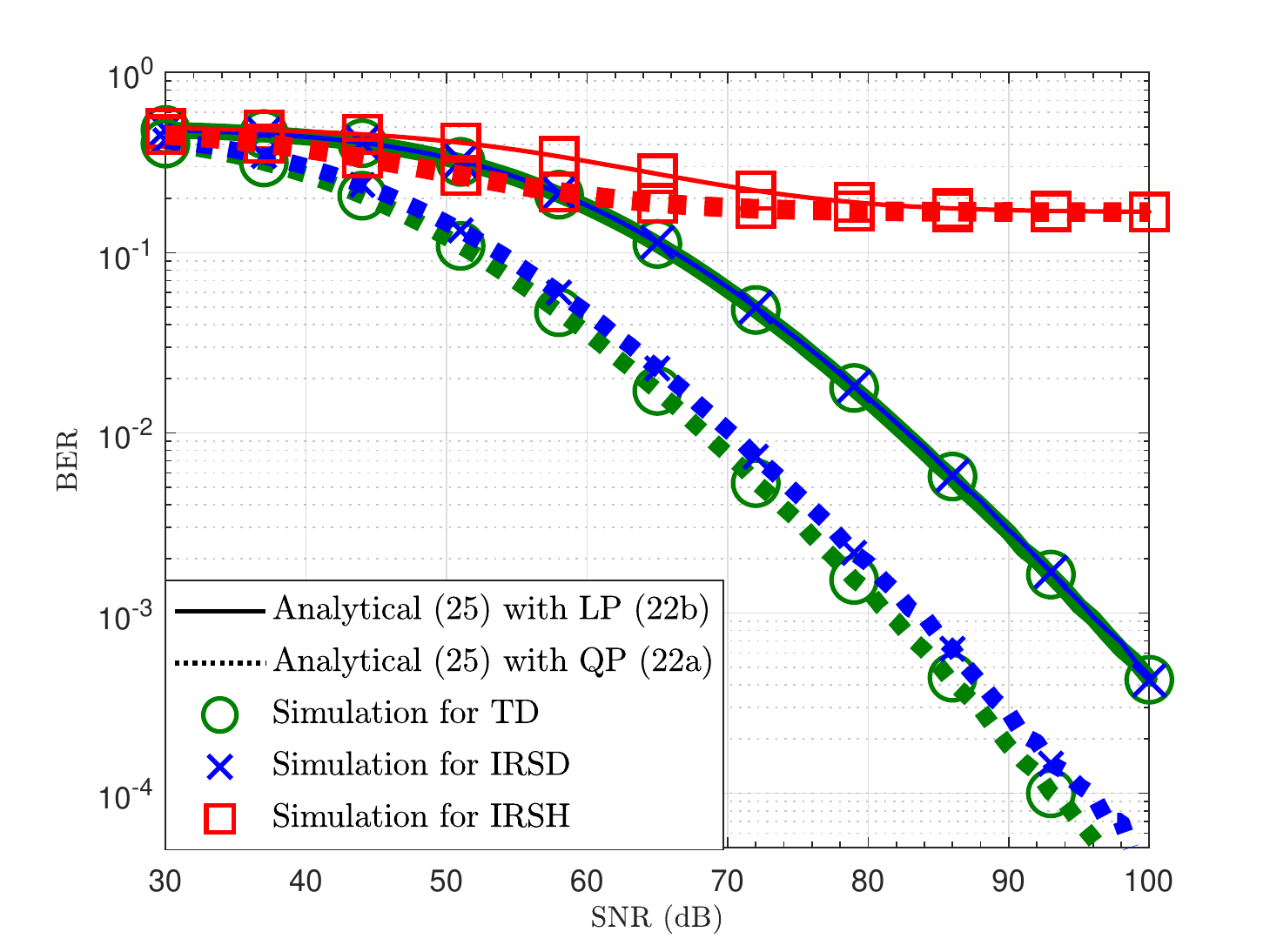}
		\vspace{-5mm}
		\caption{\label{Fig:BER1}} 		
	\end{subfigure}
	\begin{subfigure}[h]{0.49\textwidth}
		\centering
		\includegraphics[width=1\textwidth]{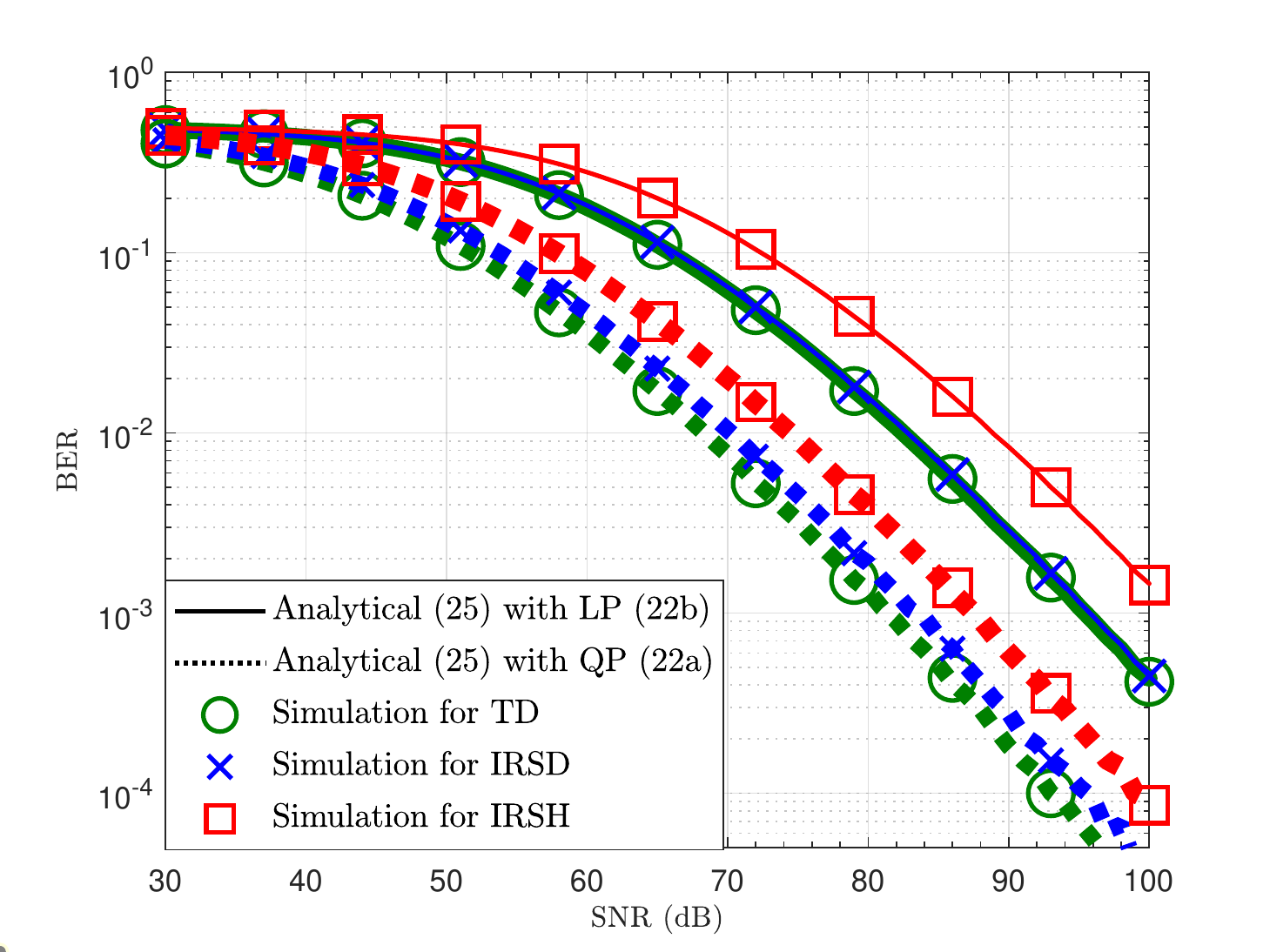}
		\vspace{-5mm}
		\caption{\label{Fig:BER2}}
	\end{subfigure}
	\vspace{-2mm}
	\caption{ BER performance versus SNR (dB) with LP and QP  profiles for three  IRS sharing protocols,  where  $\theta_{p1}=\frac{\pi}{3}$,   $\theta_{p2}=\frac{\pi}{6}$, and (a) $\Delta\theta_\ell=0$ rad and  (b) $\Delta\theta_\ell=1$ mrad.\label{Fig:BER} } \vspace{-6mm}
\end{figure}

Fig.~\ref{Fig:BER} shows  simulation  and analytical results (\ref{BER}) for the BER of the LS 1-PD 1 link for  different IRS sharing protocols employing the LP and QP profiles given in (\ref{LP}) and (\ref{QP}), respectively. Figs.~\ref{Fig:BER1} and \ref{Fig:BER2} show the BER for co-located LSs and separated LSs with $\Delta\theta_\ell=1$ mrad, respectively.  In both subfigures, the simulation results and analytical results  (\ref{BER}) match  perfectly. The TD and IRSD protocols achieve a lower BER than  the IRSH protocol. This can be explained as follows. For the considered protocols, the beam of LS 1 incident  on the IRS has beamwidths $w_x(\hat{d}_{\ell 1})=2.28$ m and $w_y(\hat{d}_{\ell 1})=1.97$ m. The tile sizes for the IRSD and IRSH protocols are $L_x=L_y=0.5$ m and $L_x=0.125$ m, $L_y=0.5$ m,  respectively, whereas the TD protocol allocates the entire IRS surface to one LS-PD pair. 
As we observed in Figs.~\ref{Fig:IntfD0} and \ref{Fig:IntfD1}, for the TD and IRSD protocols, the lens can collect more power  than for the IRSH  protocol which leads to a better BER performance in Figs.~\ref{Fig:BER1} and \ref{Fig:BER2}. Furthermore, in Fig.~\ref{Fig:BER1}, we observe a large BER degradation for the IRSH protocol. This is due to the interference caused by LS 2 at PD 1 when the LSs are co-located, see Fig.~\ref{Fig:IntfD0}.  This unfavorable behavior can be easily mitigated by separating the LSs by 1 mrad as shown in Fig.~\ref{Fig:BER2}.   Furthermore, we observe that the QP design achieves a substantial gain of up to  12 dB  over the LP design  for all  considered protocols.  The QP design in (\ref{QP})  reduces the beam divergence  along the propagation path which in turn increases the received power at the lens.  Moreover,  the TD protocol yields a 2 dB gain over the IRSD protocol for the QP profile, whereas both protocols perform similarly for the LP profile. For the LP profile, the beam diverges and since the lens is smaller than the beam footprint in the lens plane,  the performances of both protocols are similar. On the other hand, for the QP profile, the larger IRS tile for the TD protocol allows focusing more power on the lens, which leads to a better BER performance.
 
Figs.~\ref{Fig:Outage1} and \ref{Fig:Outage2} show the upper bounds on the outage probability  of the link between LS 1 and PD 1 for threshold rates $R_1=1.7$ $\mathrm{Gbit}/\mathrm{s}$ and $0.5$ $\mathrm{Gbit}/\mathrm{s}$, respectively. The  performances of the three  IRS sharing protocols are compared for misalignment  errors $|\mathbf{r}_{\ell1}-\mathbf{r}_{p1}|=0$ m and $0.17$ m. The LS locations are as specified in Table \ref{Table:Sys_Param}. In both subfigures, the simulation results perfectly match the analytical results in (\ref{Pout_last_general}).  Moreover,  by increasing  angle $\theta_{p1}$, the IRS and lens become more parallel to each other which leads to a larger   received power at  PD 1  and a  lower  outage probability. For both threshold rates, in the absence of misalignment errors, the  IRSD protocol  performs better than the IRSH and TD protocols because of the larger received power at the lens and the simultaneous transmission of both LSs, respectively. Furthermore, for the larger threshold rate, i.e., $R_1=1.7$ $\mathrm{Gbit}/\mathrm{s}$, considered in Fig.~\ref{Fig:Outage1} the  IRSH protocol performs better than the TD  protocol. This is expected since for the  TD protocol, LS 1 can transmit only in  half of the time slots. However, for the smaller threshold rate, i.e., $R_1=0.5$ $\mathrm{Gbit}/\mathrm{s}$, considered in Fig.~\ref{Fig:Outage2}, the  TD protocol  performs better than the IRSH  protocol because the benefit of  a larger received power outweighs the rate loss associated with the TD protocol in this case.  Moreover,  for both considered threshold rates, the performance of the IRSD protocol  degrades significantly for a misalignment error of $0.17$ m. This is due to the shift of the  beam of LS 1 incident on the IRS towards the tile that is configured for LS 2. This tile redirects the power  of LS 1  in the wrong direction. In contrast,  the IRSH protocol is  robust to misalignment errors, due to the homogenization of the IRS. 
\begin{figure}[t!]
	\centering
	\begin{subfigure}[h]{0.48\textwidth}
		\centering
		\includegraphics[width=1\textwidth]{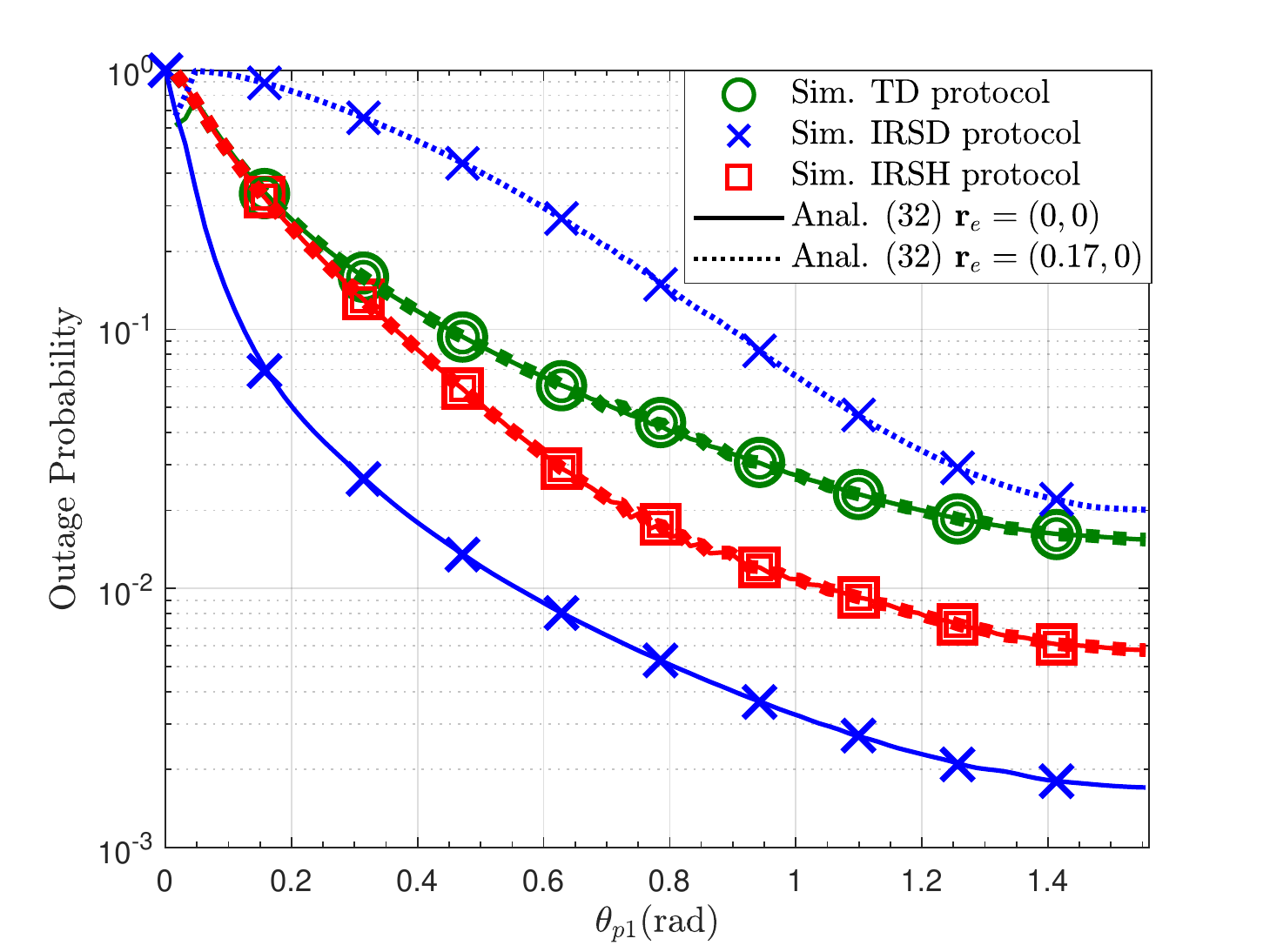}
		\vspace{-5mm}
		\caption{\label{Fig:BER1}} 		
	\end{subfigure}
	\begin{subfigure}[h]{0.49\textwidth}
		\centering
		\includegraphics[width=1\textwidth]{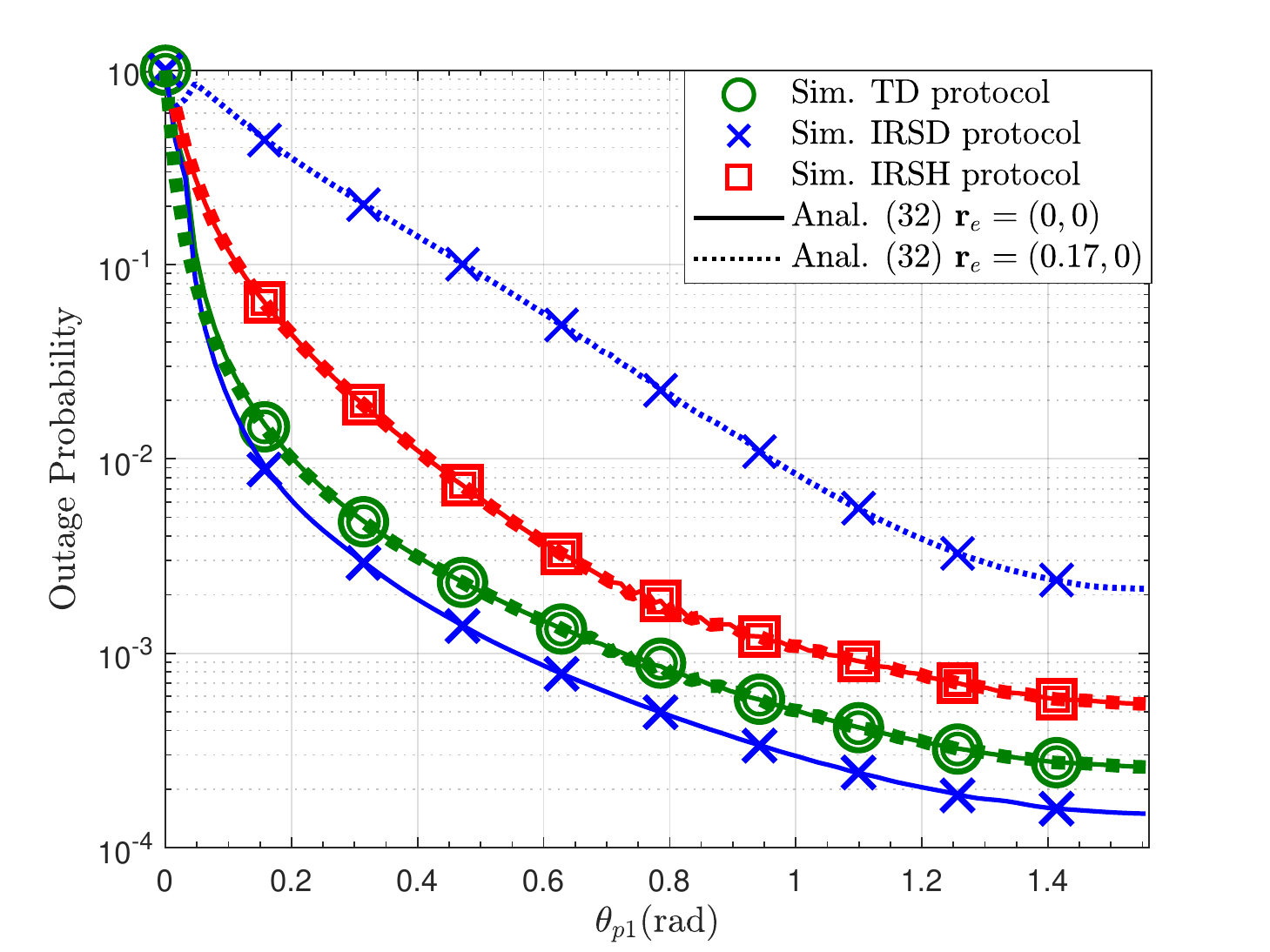}
		\vspace{-5mm}
		\caption{\label{Fig:BER2}}
	\end{subfigure}
	\vspace{-2mm}
	\caption{Upper bounds on the outage probability versus $\theta_{p1}$ (rad)  for misalignment  errors $|\mathbf{r}_e|=|\mathbf{r}_{\ell1}-\mathbf{r}_{p1}|=0$ m and $0.17$ m and threshold rates (a) $R_1=1.7$ $\mathrm{Gbit}/\mathrm{s}$ and (b) $R_1=0.5$ $\mathrm{Gbit}/\mathrm{s}$.\label{Fig:Outage}}\vspace{-8mm}
\end{figure}

\section{Conclusions}\label{Sec_concl}
In this paper, we developed  an analytical  channel model for 	point-to-point IRS-assisted FSO systems based on the Huygens-Fresnel principle. We determined the reflected electric field and the channel gain   taking into account  the non-uniform power distribution  of   Gaussian beams, the IRS size, the positions of the LS, the IRS, and the lens, and the phase shift profile of the IRS. We validated the accuracy of the proposed analytical model via simulations and showed that,  in contrast to  models based on the far-field approximation, the proposed model is valid even for intermediate distances, which are relevant in practice. Moreover, we exploited the proposed model to study the performance of an IRS-assisted  multi-link FSO system, where we proposed  three  IRS sharing  protocols. For each protocol, we designed the  size and phase shift profile of the IRS tiles, the position of the beam footprint, and the lens w.r.t. the IRS. Then, we analyzed the BER and outage probability and exploited this analysis to compare the IRS sharing  protocols in the presence of misalignment errors and  inter-link interference and for LP/QP profiles. Our results revealed that  the impact of interference can be mitigated by careful positioning of the LSs and PDs. Moreover, the  QP profile was shown to outperform the LP profile. Furthermore, which of  the proposed IRS sharing protocols is preferable, depends on the position of the LSs and PDs, the target transmission rate, and on whether or not  misalignment errors are present.
\appendices
\renewcommand{\thesectiondis}[2]{\Alph{section}:}
\vspace*{-5mm}
\section{Proof of Lemma~\ref{Lemma1}}\label{App0}
First, we express  the LS plane coordinates  $\mathbf{r}_l$ in terms of the  Cartesian coordinates $\mathbf{r}$ as  follows
\begin{IEEEeqnarray}{rll}
 {\mathbf{r}}_l=\mathbf{R}^T_{y_l}\left({\frac{\pi}{2}}-\theta_l\right) \left({\mathbf{r}}-{\mathbf{r}}_{\ell 0}\right)+\mathbf{c}_1,
\end{IEEEeqnarray}
 where $\mathbf{c}_1=(0,0,d_l)$, $\mathbf{R}_{y_l}(\cdot)$ is a rotation matrix, and ${z}=0$ since  the tile plane is located in the $xy$-plane. Then, assuming $d_l\gg L_{x}, L_{y}$, we can approximate ${z}_l\approx \hat{d}_l$ in the terms ${w}^2(\cdot)$, $R(\cdot)$, and $\tan^{-1}\left(\cdot\right)$  in (\ref{Gauss}).    Then, by substituting these variables in (\ref{Gauss}), we obtain (\ref{lem1}). 
Then, because of the law of  energy conservation  \cite{Caltech_Optic},  the  LS beam power  and the  power collected by an IRS with infinitely large area, i.e., $L_x,L_y\to \infty$,   must be the same. Thus, using (\ref{irradiance}) and (\ref{gain}), we obtain
 \begin{IEEEeqnarray}{rll}
 	\frac{1}{2\eta}	\int\limits_{-\infty}^{\infty}\int\limits_{-\infty}^{\infty}	|E_\ell(\mathbf{r}_\ell)|^2 \mathrm{d}{x}_\ell\mathrm{d}{y}_\ell=\frac{1}{2\eta}\int\limits_{-\infty}^{\infty}\int\limits_{-\infty}^{\infty} |{E}_\text{in}({\mathbf{r}})|^2 \mathrm{d}x\mathrm{d}y.
 \end{IEEEeqnarray}
 Thus, using \cite[Eq.~3.321-2]{integral}, we obtain $\zeta_\text{in}=\sqrt{\sin(\theta_\ell)}$ and this completes the proof.
\vspace*{-5mm}
 \section{Proof of Theorem~\ref{Theorem1}}\label{App1}
First, we substitute (\ref{lem1}) and $\Phi_{q}(\mathbf{r},\mathbf{r}_q^t)=\Phi_{q}^\text{quad}(\mathbf{r},\mathbf{r}_q^t)$ into (\ref{Huygens-Fresnel})  and approximate $|\mathbf{r}_o-{\mathbf{r}}|\approx\text{t}_1+\text{t}_2$, see (\ref{bionom}).  This leads to
\begin{IEEEeqnarray}{rll}
	E_q(\mathbf{r}_{o})&=  \int_{-\frac{L_y}{2}+y_q}^{\frac{L_y}{2}+y_q}  \!\!\int_{-\frac{L_x}{2}+x_q}^{\frac{L_x}{2}+x_q} A({x},{y}) \exp\left(j\Phi_\text{tot}({x},{y})\right) \mathrm{d}{x}\mathrm{d}{y},	
	\label{proofTheo1_1}
\end{IEEEeqnarray}
where  amplitude $A({x},{y})=\frac{E_0w_0\zeta_q}{\lambda{w}(\hat{d}_\ell)|\mathbf{r}_{o}-{\mathbf{r}}|}\exp\left(-{\hat{x}^2\over {w}_x^2(\hat{d}_\ell)}-{{\hat{y}}^2\over {w}_y^2(\hat{d}_\ell)}\right) $ and the total phase is the summation of the incident  beam phase, $\psi_{\text{in}}({\mathbf{r}})$, the quadratic phase shift of the tile, $\Phi_{q}^\text{quad}(\mathbf{r},\mathbf{r}_q^t)$,  and the additional  phase from the IRS to the lens, $k|\mathbf{r}_{o}-{\mathbf{r}}|$, and  is given by
\begin{IEEEeqnarray}{rll}
	\Phi_\text{tot}({\mathbf{r}})&=-\psi_{\text{in}}({\mathbf{r}})+k|\mathbf{r}_{o}-{\mathbf{r}}|-\Phi_{q}^\text{quad}(\mathbf{r},\mathbf{r}_q^t),
	\label{phas-tot}
\end{IEEEeqnarray}
where $\psi_{\text{in}}({\mathbf{r}})$ and  $|\mathbf{r}_{o}-{\mathbf{r}}|$ are respectively given by (\ref{lem1}) and approximated by terms $\text{t}_1$ and $\text{t}_2$ in (\ref{bionom}). Thus, by further approximating $|\mathbf{r}_{o}-{\mathbf{r}}|\approx |\mathbf{r}_o|$ in $A(x,y)$, we obtain
\begin{IEEEeqnarray}{rll}
	E_q(\mathbf{r}_{o})&=\tilde{C} \int_{-\frac{L_x}{2}+x_q}^{\frac{L_x}{2}+x_q} e^{-\tilde{b}_x{{x}^2}-jkx\left(\tilde{A}_0+\Phi_{q,x}-2x_q^t\Phi_{q,x^2}\right)} \mathrm{d}{x} \int_{-\frac{L_y}{2}+y_q}^{\frac{L_y}{2}+y_q}  e^{-\tilde{b}_y{{y}^2}-jky\left(\tilde{B}_0+\Phi_{q,y}-2y_q^t\Phi_{q,y^2}\right)} \mathrm{d}{y}.	\quad
	\label{seperatexy}
\end{IEEEeqnarray}
where $\tilde{C}=\frac{E_0w_0\zeta_q\zeta_\text{in}}{j\lambda w(\hat{d}_\ell) |\mathbf{r}_o|}e^{-\nu \sin^2(\theta_\ell) x_{\ell 0}^2-\nu y_{\ell 0}^2+j\tan^{-1}(\frac{\hat{d}_\ell}{z_0})+jk\left(|\mathbf{r}_o|-\hat{d}_\ell+\delta_q\right)}$,  $\tilde{b}_x=\nu\sin^2(\theta_\ell)-\frac{jk}{2|\mathbf{r}_o|}\left(1-\frac{x_o^2}{|\mathbf{r}_o|^2}\right)+jk\Phi_{x^2}$, 
$\tilde{b}_y=\nu-\frac{jk}{2|\mathbf{r}_o|}\left(1-\frac{y_o^2}{|\mathbf{r}_o|^2}\right)+jk\Phi_{y^2}$, $\tilde{A}_0={2j\nu{x}_{\ell0}\over k}\sin^2(\theta_\ell)+\frac{x_o}{|\mathbf{r}_o|}-\cos(\theta_\ell)$, and $\tilde{B}_0={2j\nu{y}_{\ell0}\over k}+\frac{y_o}{|\mathbf{r}_o|}$.
Next,  we apply  \cite[Eq.~(2.33-1)]{integral} 
\begin{IEEEeqnarray}{rll}
	\int e^{-a{x^2}-bx} \mathrm{d}x=\frac{1}{2}\sqrt{\frac{\pi}{a}}\exp\left(\frac{b^2}{4a}\right)\text{erf}\left(\sqrt{a}x+\frac{b}{2\sqrt{a}}\right),
\end{IEEEeqnarray}
to solve the integrals in (\ref{seperatexy}). Then, we express $\mathbf{r}_o$ in terms of $\mathbf{r}_p$ using the following relation 
\begin{IEEEeqnarray}{rll}
	{\mathbf{r}_o}=\mathbf{R}_z(-{\phi_p})\mathbf{R}_{y_p}(\frac{\pi}{2}-{\theta_p})\left({\mathbf{r}_p}+\mathbf{c}_2\right)+{\mathbf{r}}_{p0},
	\label{ro-rp}
\end{IEEEeqnarray}
where $\mathbf{c}_2=(0, 0, d_p)$ is the translation vector  and $\mathbf{R}_{y_p}(\cdot)$ and $\mathbf{R}_z(\cdot)$ are rotation matrices, respectively. Now,  we substitute $x_o$, $y_o$ from (\ref{ro-rp}) in the integral results of (\ref{seperatexy}) and approximate $|\mathbf{r}_o|\approx d_p$, $z_o\approx d_p\sin(\theta_p)$, $\frac{x_o^2}{|\mathbf{r}_o|^3}=c_5^2d_p+\frac{2}{d_p}c_5x_{p0}$, and $\frac{y_o^2}{|\mathbf{r}_o|^3}=c_6^2d_p+\frac{2}{d_p}c_6y_{p0}$. This leads to (\ref{theo1}) and  completes the proof.
 \vspace*{-5mm}
 \section{Proof of Theorem~\ref{Theorem4}}\label{App4}
First, after  substituting (\ref{theo1}) into (\ref{Final-gain}), we obtain
 \begin{IEEEeqnarray}{rll}
 	h_{\text{irs}}=\frac{2}{\pi E_0^2w_0^2}\sum_{q=1}^Q \sum_{\varsigma=1}^Q\iint_{\mathbf{r}_p\in \mathcal{A}_p}  E_q(\mathbf{r}_p)  \left(E_\varsigma(\mathbf{r}_p)\right)^* \mathrm{d}\mathbf{r}_p.
  \end{IEEEeqnarray}
Since  for the size of the lens, $\frac{x_p}{d_p}, \frac{y_p}{d_p}\ll 1$ holds, we approximate $x_p=y_p\approx \frac{a}{2}$ in the  $\text{erf}(\cdot)$  terms in  (\ref{theo1})  and substitute   $C_{2,q}$ and $C_{2,\varsigma}$. Then, we obtain
 \begin{IEEEeqnarray}{rll}
 	h_{\text{irs}}&=C_h\sum_{q=1}^Q \sum_{\varsigma=1}^Q C_q C_\varsigma^* C_{2,q} C_{2,\varsigma}^*   \int\limits_{-a}^{a}\!\int\limits_{-\epsilon}^{\epsilon}   \!\exp\!\Bigg(\!\!-\frac{k^2}{4}\Big(\frac{1}{{b}_x}\left(A_0+c_1x_p+c_2y_p+\Phi_{q,x}-2x_q^t\Phi_{q,x^2}\right)^2\nonumber\\
 	&+\frac{1}{{b}_y}\!\!\left(B_0+c_3x_p+c_4y_p+\Phi_{q,y}-2y_q^t\Phi_{q,y^2}\right)^2\!\!+\frac{1}{{b}_x^*}\left({A}_0^*+c_1x_p+c_2y_p+\Phi_{\varsigma,x}-2x_\varsigma^t\Phi_{\varsigma,x^2}\right)^2\!\!\!\nonumber\\
 	&+\frac{1}{{b}_y^*}\!\!\left(B_0^*+c_3x_p+c_4y_p+\Phi_{\varsigma,y}-2y_\varsigma^t\Phi_{\varsigma,y^2}\right)^2\!\Big)\!\!\Bigg)\mathrm{d}y_p \,\mathrm{d}x_p,\!\!\!\quad
 	\label{ptheo42}
  \end{IEEEeqnarray}
where $\epsilon=\sqrt{a^2-x_p^2}$. Next, given the small size of the circular lens, it can be  approximated   by  a  square lens with the same area and length $a\sqrt{\pi}$. Thus, we can rewrite (\ref{ptheo42}) as follows
  \begin{IEEEeqnarray}{rll}
 	h_{\text{irs}}&={C}_h\sum_{q=1}^Q \sum_{\varsigma=1}^Q C_q C_\varsigma^* C_{2,q} C_{2,\varsigma}^* \int\limits_{-\tilde{a}}^{\tilde{a}}\int\limits_{-\tilde{a}}^{\tilde{a}}   e^{-\left(\rho_xx_p^2+\rho_yy_p^2+\rho_{xy}x_p y_p+\varrho_xx_p+\varrho_yy_p\right)}\mathrm{d}x_p \, \mathrm{d}y_p.\!\qquad
 	\label{ptheo43}
 \end{IEEEeqnarray}
Then, we solve the inner integral by applying \cite[Eq.~(2.33-1)]{integral}, which leads to (\ref{theo4})  and  completes the proof.
\vspace*{-5mm}
 \section{Proof of Proposition~\ref{Passivity}}\label{App5}
First,  we assume a tile with size $L_x, L_y\to \infty$ and a lens with radius $a\to \infty$ to determine  the maximum  powers received  by the tile and the lens. Moreover, let us assume $\mathbf{r}_{0\ell}=\mathbf{r}_{0p}=\mathbf{r}_{q}=\mathbf{r}_q^t=\mathbf{0}$. Then, given the passivity of the  tile, the total average power on its surface should be zero, or equivalently, the reflected power and the incident power are equal.  Thus, we obtain $P_\text{irs}=P_\text{in}$, where $P_\text{irs}$ is given by (\ref{irradiance}) and (\ref{theo1}) as follows
\begin{IEEEeqnarray}{rll}
P_\text{irs}=\int\limits_{-\infty}^{\infty}\int\limits_{-\infty}^{\infty}  I^{m,n}_\text{irs}(\mathbf{r}_{pn})\, \mathrm{d}x_{pn}\mathrm{d}y_{pn}\overset{(a)}{=} \frac{P_{\ell m} \zeta_0^2\bar{\zeta}_q^2{\zeta}_\text{in}^2}{|\sin(\theta_{pn})||\sin(\theta_{\ell m})|},
\label{proof_prop1}
\end{IEEEeqnarray}
where in $(a)$  we apply  \cite[Eq.~(3.323-2), Eq.~(3.321-3)]{integral} and the incident power is given by
\begin{IEEEeqnarray}{rll}
P_\text{in}=\int\limits_{-\infty}^{\infty}\int\limits_{-\infty}^{\infty} I_\text{in}(\mathbf{r})\mathrm{d}x\mathrm{d}y=\frac{1}{2\eta} \int\limits_{-\infty}^{\infty}\int\limits_{-\infty}^{\infty} \lvert E_\text{in}(\mathbf{r})\rvert^2\mathrm{d}x\mathrm{d}y=P_{\ell m},
\label{proof_prop2}
\end{IEEEeqnarray}	
 where $E_\text{in}(\mathbf{r})$ is given in (\ref{lem1}). Then,  (\ref{proof_prop1}) and (\ref{proof_prop2}) have to be equal and by assuming $\zeta_0=1$ and substituting  $\zeta_\text{in}=\sqrt{\lvert\sin(\theta_{\ell m})\rvert}$ from (\ref{lem1}),  we obtain $\bar{\zeta}_q$  in (\ref{zeta}) and this completes the proof.
 \vspace*{-5mm}
 \section{Proof of Lemma~\ref{remk_BER}}\label{App6}
 An  error  occurs when  the $n$-th PD cannot correctly detect the OOK symbol, $s_n$, which is transmitted by the $n$-th LS. Assuming equally probable OOK modulated symbols $s_n\in\{0,1\}$, we obtain
\begin{IEEEeqnarray}{rll}
&\text{Pr}\left(s_n\neq \hat{s}_n\right)=\frac{1}{2^N}\left[\text{Pr}\left(s_n\neq \hat{s}_n| s_n=0, \mathbf{s}_m\in \mathcal{S}_m\right)+\text{Pr}\left(s_n\neq \hat{s}_n|s_n=1,\mathbf{s}_m\in\mathcal{S}_m\right)\right],\quad m\neq n,\qquad\nonumber
\end{IEEEeqnarray}
where $\mathcal{S}_m$ denotes the set of all  $2^{N-1}$ possible vectors of interfering symbols.
Then, considering that the Gaussian noise $w_n$ follows distribution $f_n(w_n)=\frac{1}{\sqrt{2\pi}\sigma_n}\exp\left(-\frac{n^2}{2\sigma_w^2}\right)$, the BER is given by
 \begin{IEEEeqnarray}{rll}
 &\text{Pr}\left(s_n\neq \hat{s}_n\right)=\frac{1}{2^{N}}\Bigg[\sum_{\mathbf{s}_m\in\mathcal{S}_m}\int_{h_a^{n,n} \frac{\sqrt{\gamma_{n}}}{2}}^{\infty} f_n\left(y_n-\sum_{\substack{m\in\mathcal{N}\\ m\neq n}} h_a^{m,n} s_m \sqrt{\gamma_{m}}\right)\, \mathrm{d}y_n+\nonumber\\
&\sum_{\mathbf{s}_m\in\mathcal{S}_m} \int_{-\infty}^{h_a^{n,n}\frac{ \sqrt{\gamma_{n}}}{2}} f_n\left(y_n-h_a^{n,n} \sqrt{\gamma_n}-\sum_{\substack{m\in\mathcal{N}\\ m\neq n}}h_a^{m,n} s_m\sqrt{\gamma_{m}}\right)\, \mathrm{d}y_n \Bigg].\qquad
\end{IEEEeqnarray}
Next, using the definition of the Gaussian Q-function,  $Q(x)=\frac{1}{\sqrt{2\pi}}\int_{x}^{\infty}e^{-{u^2\over2}} \mathrm{d}u$, leads to (\ref{PEP}) and this completes the proof.
\vspace*{-5mm}
 \section{Proof of Lemma~\ref{remk_CAP}}\label{App7}
 From \cite{Lapid} it is known that the capacity achieving input distribution for FSO systems is non-Gaussian. Thus, we assume $s_m[t],\forall m\in\mathcal{N}$, is non-Gaussian. Then, assuming  Gaussian distributed noise $w_n[t]$, we can rewrite the system model in (\ref{system_eq}) as follows
\begin{IEEEeqnarray}{rll}
 y_n[t]=h_{n,n}s_n[t]+\tilde{w}_n[t],	
 \label{sys_proof}
\end{IEEEeqnarray}
where $\tilde{w}_n[t]=\sum\limits_{ m\neq n}h_{m,n}s_m[t]+w_n[t]$ is an additive non-Gaussian noise with variance $\tilde{\sigma}_{w}^2=\lvert\sum\limits_{ m\neq n}h_{m,n}\rvert^2$ $P_{\ell m}+\sigma_w^2$.  To provide a lower bound for  the capacity of the system in (\ref{sys_proof}), we substitute the above system by a system with Gaussian noise
\begin{IEEEeqnarray}{rll}
	y_n[t]=h_{n,n}s_n[t]+\hat{w}_n[t],	
	\label{sys_Gaus_proof}
\end{IEEEeqnarray}
where $\hat{w}_n[t]$ is Gaussian noise with the same variance as $\tilde{w}_n[t]$, i.e., $\hat{\sigma}_{{w}}^2=\tilde{\sigma}_{w}^2$. Then,  according to \cite[Eq.~(4)]{Non-Gauss-Cap},  the capacity of the non-Gaussian noise system in (\ref{sys_proof}), ${\mathcal{C}}_n$, is lower bounded by the capacity of the equivalent system with Gaussian noise in (\ref{sys_Gaus_proof}), $\hat{\mathcal{C}}_n$,  as follows
\begin{IEEEeqnarray}{rll}
	\hat{\mathcal{C}}_n\leq {	\mathcal{C}_n}.
	\label{Lowerband1}
\end{IEEEeqnarray}
 Assuming  an average power constraint, $\mathbb{E}\{|s_n[t]|^2\}\leq P_{\ell n}$, we can achieve a tight lower bound for the capacity of the Gaussian channel in (\ref{sys_Gaus_proof}) by adopting  exponentially distributed input symbols \cite{Lapid}. The lower bound is given by  
 \begin{IEEEeqnarray}{rll}
\hat{\mathcal{C}}_n\geq	\mathcal{C}_{\text{low}}(\Upsilon_n)=\frac{W_\text{FSO}}{2} \ln \left(1+\frac{e|h_{n,n}|^2P_{\ell n}}{2 \pi\hat{\sigma}_{{w}}^2}\right).
 	\label{Lower_proof_formula}
 \end{IEEEeqnarray} 
Thus, using the relation in (\ref{Lowerband1}) and the lower bound in (\ref{Lower_proof_formula}), we obtain 
 \begin{IEEEeqnarray}{rll}
 	\mathcal{C}_n\geq \hat{	\mathcal{C}}_n\geq \mathcal{C}_{\text{low}}(\Upsilon_n),
 		\label{Loweband2}
 \end{IEEEeqnarray}	
 and thus, $\mathcal{C}_{\text{low}}(\Upsilon_n)$ is a lower bound for the capacity of the system in (\ref{system_eq}). Substituting the value of $\hat{\sigma}_{{w}}$ by $\tilde{\sigma}_{w}$ in (\ref{Lower_proof_formula}),  leads to (\ref{Interf_Cap}) and   completes the proof.
\bibliographystyle{IEEEtran}
\bibliography{My_Citation_1-07-2020}
\end{document}